\documentclass[11pt]{article}
\usepackage[a4paper, margin = 1in]{geometry}
\usepackage[sorting = anyt, maxbibnames=9,maxcitenames=2, backend = bibtex, style=authoryear, doi=false, url=false, dashed=false]{biblatex}
\usepackage[titletoc,page]{appendix}
\renewcommand*\appendixpagename{Appendix}
\renewcommand*\appendixtocname{Appendix}

\renewbibmacro{in:}{}
\usepackage{graphicx}
\usepackage{setspace}
\usepackage{gensymb}
\usepackage{enumerate}
\usepackage{abstract,lipsum}
\usepackage[colorlinks=true,linkcolor=blue,citecolor=blue,urlcolor=blue]{hyperref}

\usepackage{amsmath}
\usepackage{amssymb}
\usepackage{amsthm}
\usepackage{amsfonts}
\usepackage{upgreek}
\usepackage{mathtools}
\usepackage{mathrsfs}
\usepackage{bm}
\newcommand\defeq{\stackrel{\mathclap{\normalfont\tiny\mbox{def}}}{=}}
\usepackage[symbol]{footmisc}

\usepackage{color}
\usepackage{multirow}
\usepackage{algorithm,algorithmic}
\usepackage{verbatim}

\usepackage{tcolorbox}
\usepackage[english]{babel}
\theoremstyle{remark}

\begin{document}

%\tableofcontents
%\clearpage

\begin{center}
       \fontsize{15pt}{15pt}\selectfont \textbf{Large strain micromechanics of thermoplastic elastomers with random microstructures}
       
       \vspace*{0.3in}
       \fontsize{10}{10pt}\selectfont Hansohl Cho$^{1,\dagger}$, Jaehee Lee$^{1}$, Jehoon Moon$^{1}$, Elmar P\"{o}selt$^{2}$, Pieter J. in 't Veld$^{3}$, \\
       Gregory C. Rutledge$^{4,\ddagger}$, Mary C. Boyce$^{5,*}$ \\       
       \vspace*{0.3in}
       \fontsize{9pt}{9pt}\selectfont $^{1}$Department of Aerospace Engineering, Korea Advanced Institute of Science and Technology, Daejeon 34141, Republic of Korea \\ 
       \vspace*{0.1in}       
       \fontsize{9pt}{9pt}\selectfont $^{2}$BASF Polyurethane GmbH, Lemf\"{o}rde 49448, Germany  \\
       \vspace*{0.1in}    
       \fontsize{9pt}{9pt}\selectfont $^{3}$BASF SE, Ludwigshafen 67056, Germany \\
       \vspace*{0.1in} 
       \fontsize{9pt}{9pt}\selectfont $^{4}$Department of Chemical Engineering, Massachusetts Institute of Technology, Cambridge, MA 02139, US  \\
       \vspace*{0.1in}
       \fontsize{9pt}{9pt}\selectfont $^{5}$Department of Mechanical Engineering, Columbia University, New York, NY 10027, US  \\

\vspace*{0.2in}
\fontsize{8.5pt}{8.5pt}\selectfont E-mails: $^\dagger$hansohl@kaist.ac.kr (H. Cho); $^\ddagger$rutledge@mit.edu (G. C. Rutledge); $^{*}$boyce@columbia.edu (M. C. Boyce)

%\vspace*{0.2in}
%\fontsize{10pt}{10pt}\selectfont Date: April 6, 2023
\end{center}

\renewenvironment{abstract}
{\small 
\noindent \rule{\linewidth}{.5pt}\par{\noindent \bfseries \abstractname.}}
{\medskip\noindent \rule{\linewidth}{.5pt}
}

\vspace*{0.3in}
\onehalfspacing
\begin{abstract}
\fontsize{10pt}{10pt}\selectfont
Thermoplastic polyurethanes (TPU) are block copolymeric materials composed of plastomeric “hard” and elastomeric “soft” domains, by which they exhibit highly resilient yet dissipative large deformation features depending on volume fractions and microstructures of the two distinct domains. Here, we develop a new methodology to address the microscopic deformation mechanisms in TPU materials with highly disordered microstructures. We propose new micromechanical models for randomly dispersed (or occluded) as well as randomly continuous hard domains, each within a continuous soft structure as widely found in representative TPU materials over a wide range of volume fractions, v$_{\mathrm{hard}}$ = 26.9\% to 52.2\%. The micromechanical modeling results are compared to experimental data on the macroscopic large strain behaviors reported previously (\cite{cho2017deformation}). We explore the role of the dispersed vs. continuous nature of the geometric features of the random microstructures on shape recovery and energy dissipation at the microstructural level in this important class of phase-separated copolymeric materials. \\ \\
% This paper addresses the microstructural deformation mechanisms in thermoplastic polyureth-ane (TPU) materials and their connections to the macroscopic deformation features reported in our previous paper (\cite{cho2017deformation}). Toward this end, we propose new three-dimensional representative volume elements of continuous or dispersed domains that possess key geometric features of microstructures often found in the TPU materials and other two-phase heterogeneous materials by employing random spatial tessellation. Furthermore, we present a systematic procedure to elucidate the key micro-mechanisms involving elasticity, plasticity, deformation-induced softening also known as Mullins’ effect, hysteresis and shape recovery in this important class of phase-separated elastomeric (or plastomeric) materials at the microstructural level. \\ \\
\end{abstract}

\doublespacing
\section{Introduction}
Thermoplastic polyurethanes (TPU) are attractive materials for diverse engineering, defense and biological applications owing to their versatile mechanical properties involving both elasticity and inelasticity. TPU materials are often composed of alternate, distinct “hard” and “soft” contents, which are thermodynamically immiscible. Tailoring the phase separation and microstructures in TPU materials has also been shown to be flexible via synthesis methods, chemical compositions and volume fractions of the hard and soft contents in the materials; readers are referred to \cite{cho2017deformation} and references therein for more details. The presence of phase-separated morphologies comprising distinct hard and soft domains often leads to remarkably resilient yet dissipative mechanical features with a cross of elastomeric and thermoplastic polymeric properties, especially under large deformations. The geometric features of these two components have also been found to influence critically the macroscopic mechanical behavior of the materials in conjunction with the intrinsic mechanical properties as well as the volume fractions of hard and soft components. Thus, micromechanical modeling of the substructures has been widely employed to elucidate the underlying deformation mechanisms and the connections between the macroscopic mechanical behavior and the microstructures in this important class of heterogeneous soft materials.

Micromechanical modeling, through the identification of morphologically complex multi-phase domains, has become an important tool for assessing the underlying structural principles for the macroscopic deformation features in rubbery, glassy and semicrystalline polymers as well as polymer nanocomposites possessing secondary and tertiary phases or voids. Beyond the classical work by \cite{mori1973average} to predict the effective properties of linear elastic materials with inclusions, the large strain micromechanical behavior of filled elastomers has been investigated for various geometries of hard fillers (\cite{bergstrom1999mechanical, bergstrom2000large, govindjee1991micro, govindjee1992transition, govindjee1992mullins,  castaneda1996exact, castaneda2000second, lopez2006overall, lopez2013nonlinear, leonard2020nonlinear}). The large strain micromechanical behavior of thermoplastic glassy polymers and thermoplastic semicrystalline polymers has also been widely investigated (\cite{socrate2000micromechanics, socrate2001micromechanical, van2003micromechanical, boyce2001micromechanisms, boyce2001micromechanics, danielsson2002three, parsons2004experimental, danielsson2007micromechanics, tzika2000micromechanics, van2003micromechanical}), by which toughening mechanisms for relatively brittle thermoplastic materials were addressed with the introduction of a spatial distribution of second-phase particles or voids. Moreover, numerous micromechanical models have been proposed to predict the effective mechanical properties of polymer nanocomposites possessing inorganic particles or fibers with various geometric features (\cite{sheng2004multiscale, zeng2008multiscale, pukanszky2005interfaces, mortazavi2013modeling}). While there is a large amount of research on micromechanical models of rubbery, glassy and semicrystalline polymers and polymer composites with secondary inclusions or voids, micromechanical models of thermoplastic elastomers exhibiting hybrid mechanical features of rubbery and thermoplastic polymers through complex phase-separated morphologies are largely lacking. Most microscopic studies of these materials have focused on fully atomistic or coarse-grained molecular dynamics simulations (\cite{lempesis2016simulation, lempesis2017atomistic, zhu2018tens, zhu2018comp, manav2021molecular, heyden2016all, cui2013thermomechanical}) by which characteristics of the phase separation and segregation of the hard and soft components were explored and the micromechanical behaviors were quantified, especially at ultrafast strain rates such as those associated with shock loading or a ballistic impact (\cite{zhang2018mechanical, liu2019coarse, eastmond2021probing, liu2019coarse}). Moreover, a few micromechanical models at the continuum level were presented to explore the micro-mechanisms of deformation, recovery and softening in thermoplastic elastomers that exhibit highly resilient yet dissipative, large strain behaviors (\cite{boyce2001micromechanisms, boyce2001micromechanics}). However, these continuum mechanics-based micromechanical models with crude idealization of the complex morphologies of two-phase systems have been limited to assessing the microstructural mechanisms in thermoplastic elastomers with relatively high fractions of second-phase particles. Moreover, the unrealistic kinematic constraints used in the micromechanical models enable explorations of the micro-mechanisms only in two-dimensional deformation modes.

In this work, we explore microstructural deformation mechanisms in representative thermoplastic polyurethane systems with a wide range of volume fractions of hard and soft components. We propose three-dimensional representative volume elements (RVE) capable of accommodating arbitrary three-dimensional deformation modes. Specifically, the three-dimensional RVEs that mimic more realistic, highly disordered microstructures often found in TPU materials are presented, for which Voronoi-tessellation via random spatial points is employed with appropriate three-dimensional periodic boundary conditions. Then, we discuss the underlying microscopic deformation mechanisms and connections between the micro-mechanisms and the macroscopic deformation features at small to large strains in TPUs reported in our previous paper (\cite{cho2017deformation}). Lastly, the implications of the micromechanical modeling results on deformation, recovery and cyclic softening are discussed.

The paper is organized as follows. The macroscopic mechanical features for TPU materials differing in terms of the volume fractions of hard and soft components are briefly summarized in Section \ref{section2}. Representative volume elements for the TPU materials with two different morphologies, (1) dispersed hard domains and (2) their continuous counterparts, are then introduced with the micromechanical modeling framework in Section \ref{section3}. In Section \ref{section4}, we compare the micromechanical modeling results for TPU materials with corresponding experimental data in various loading scenarios, by which we further discuss the implications of the micromechanical modeling results for microscopic deformation mechanisms in these TPU materials. In the Appendices \ref{appendixA} and \ref{appendixB}, detailed information about the constitutive models used in hard and soft components in the micromechanical analysis is provided together with the corresponding computational implementation procedures. Furthermore, more detailed geometric information on the newly developed representative volume elements used throughout the main body of the manuscript is provided in the Appendix \ref{appendixC}.

\section{Macroscopic mechanical behavior of TPUs: a brief review of resilience and dissipation}
\label{section2}
\begin{figure}[b!]
\centering
\includegraphics[width=0.7\textwidth]{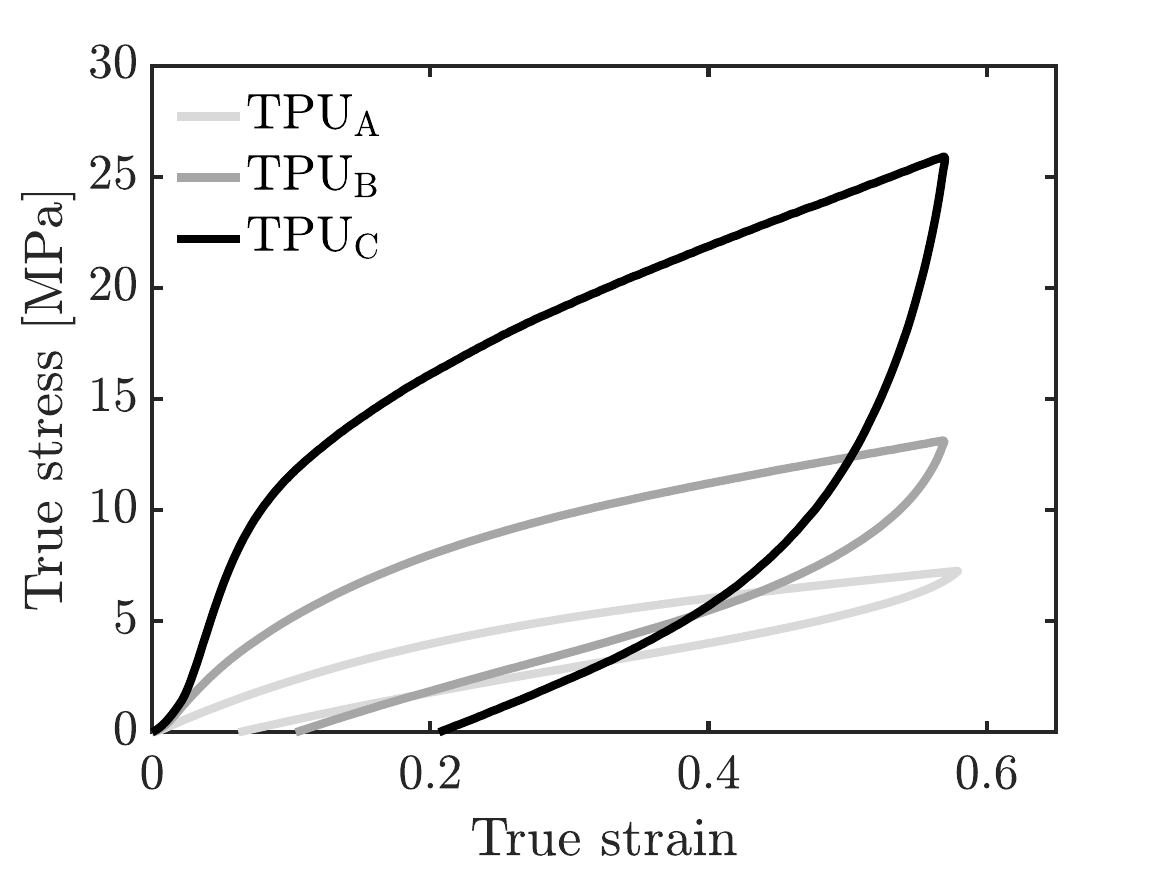} 
\caption{Stress-strain behavior of TPU materials with varying volume fractions of hard and soft components under uniaxial compression at a strain rate of 0.01 $\mathrm{s}^{-1}$; TPU$_{\mathrm{A}}$ (v$_{\mathrm{hard}}=26.9 \%$), TPU$_{\mathrm{B}}$ (v$_{\mathrm{hard}}=39.3 \%$) and TPU$_{\mathrm{C}}$ (v$_{\mathrm{hard}}=52.2 \%$). Experimental data taken from \cite{cho2017deformation}.}
\label{fig:experimental_results}
\end{figure}
\noindent
Figure \ref{fig:experimental_results} summarizes the large strain mechanical behavior of representative TPU materials differing in terms of the volume fractions of hard and soft components under cyclic loading and unloading conditions. Highly nonlinear elastic and inelastic features were observed in the experimental data, strongly depending on the volume fractions of the constituents. It is also undisputed that a TPU material with a higher fraction of hard components exhibits a greater stress response with a more apparent yield-like stress rollover accompanied by greater hysteresis. However, TPU materials with a lower fraction of hard components clearly exhibit more “rubbery” features involving less hysteresis yet remarkable elastic shape recovery upon unloading. It should also be noted that a small change in the volume fraction of hard and soft components results in a significant change in all of the large strain mechanical features, including the initial stiffness, yield, post-yield behavior, energy dissipation and elastic shape recovery upon cyclic loading conditions. More details about the macroscopic mechanical behaviors of these TPU materials can be found in our previous paper (\cite{cho2017deformation}).

\section{Representative morphology and micromechanical modeling framework}
\label{section3}
\begin{figure}[h!]
\centering
\includegraphics[width=1.0\textwidth]{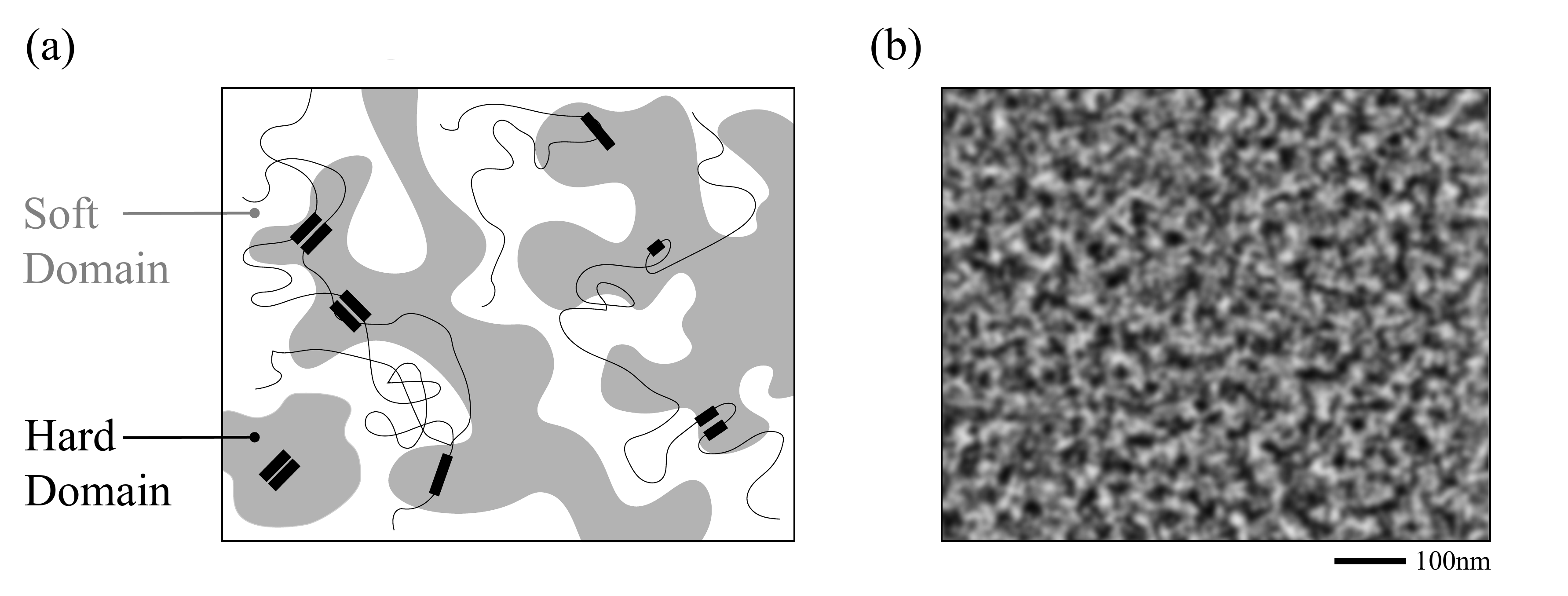} 
\vspace{-0.3in}
\caption{Microstructures and morphologies in representative TPU materials: (a) Schematic of continuous or isolated hard domains within soft matrices \textcolor{blue}{(\cite{rinaldi2011microstructure,qi2005stress,hepburn2012polyurethane})}, and (b) Transmission electron microscopy image of an exemplar TPU material used in this study, where the bright regions are hard domains while the dark regions are soft domains.}
\label{fig:schematic}
\end{figure}

Phase-separation and morphologies in TPU materials have been widely studied using microscopy and X-ray scattering measurements, by which the phase-separation and relevant geometric features were found to be strongly dependent on the thermodynamic properties and interactions of the constituents (\cite{choi2012microstructure, castagna2012role, he2014role, choi2009influence, garrett2000microphase, stribeck2017thermoplastic, stribeck2019melting}). For this study, we selected representative TPUs in which the two domains separate well with less phase-mixing. Specifically, more polarized hard domains consisting of methylene diphenyl diisocyanate (MDI) and 1,4-butane diol repeating units phase-separate better due to the polarity differences compared to soft domains, hydrogen bonding of the urethane bonds of the hard blocks and the $\pi-\pi$ stacking of the aromatic rings of MDI. Moreover, the soft domains consist mostly of polytetrahydrofuran while very small MDI blocks may be occluded inside the soft domains. Meanwhile, a wide variety of microstructures have been reported in these phase-separated TPU materials, especially depending on the volume fractions of the constituents. Specifically, the continuous hard domain microstructures within soft matrices or interpenetrating networks of co-continuous hard and soft domains are present in TPUs with a relatively high fraction of hard contents. By contrast, the hard domain microstructures are more likely to be dispersed or isolated within the soft matrices in TPUs with a relatively low fraction of hard contents. It should also be noted that the local microstructures are highly disordered for the two representative morphologies with isolated or continuous hard microstructures in the materials. The representative phase-separated morphology in TPUs is schematically presented in Figure \ref{fig:schematic}a. Moreover, Figure \ref{fig:schematic}b depicts an exemplar microscopy image of a representative TPU (one of our target TPU materials with a volume fraction of hard components of  $\sim$ 52.2\%), where the continuous hard morphology is well observed. 
% Figure2

\subsection{Construction of representative volume elements}
\label{section31}
In the studies of micromechanics of multi-phase, heterogeneous materials, the first step is to construct an appropriate representative volume element (RVE) capable of capturing the essential features of microstructural geometries as well as the various deformation modes in the materials. The RVEs of TPU materials considered in this study require the two main features: (1) a random distribution of the microstructural domains and (2) continuous or dispersed hard domains in three dimensions. Toward this end, we present simple and systematic procedures that can be used to construct highly disordered three-dimensional RVEs with continuous or dispersed hard domains. Moreover, inter-phase or phase mixing between hard and soft domains is assumed to be negligible in our study.

\subsubsection{Dispersed hard domain}
\label{section311}
Micromechanical models of particle- or void-dispersed materials have long been studied for composites and cellular materials at a broad range of length scales. Most of the classical micromechanical models have idealized the microstructures of particle or void distributions as stacked arrays of particles, which usually yields unrealistic predictions of the corresponding micromechanical behavior (\cite{tvergaard1982localization, tvergaard1982ductile, koplik1988void, steenbrink1997void}). More suitable representations of particle distributions can be obtained if the particles are staggered rather than stacked. Specifically, \cite{socrate2000micromechanics} and \cite{danielsson2002three} proposed axisymmetric and three-dimensional RVEs for particle-dispersed thermoplastic materials based on a body-centered-cubic array of particles capable of more suitably capturing the three-dimensional microstructural features of the materials. The staggered-particle approach has been recently extended to model the micromechanical behaviors of two-phase soft materials for which three-dimensional RVEs staggered on various crystalline lattices were presented (\cite{cho2016engineering}). However, the particle distributions in two-phase material systems are typically random. Hence, RVEs possessing a random distribution of particles or voids have been widely studied. \cite{smit1999prediction} proposed a single-void-based two-dimensional RVE that is able to capture the two-dimensional percolation of plastic deformation within glassy polymeric matrices. Then, much improved idealization of randomly staggered RVEs was realized in \cite{parsons2004experimental}, \cite{parsons2005three}, \cite{parsons2006mechanics}, \cite{danielsson2003micromechanics}, \cite{danielsson2007micromechanics}, \cite{jandron2018numerical}, \cite{jandron2020electromechanical}, \cite{tarantino2019random} and \cite{zerhouni2019numerically}, where several particles or voids were considered within RVEs. In these studies, systematic procedures for constructing random microstructures of multiple voids or particles with mono- or poly-dispersion in sizes were presented. Similar to the multi-void approach proposed by \cite{danielsson2007micromechanics}, we herein construct three-dimensional RVEs possessing randomly dispersed hard domains for the given volume fractions of hard components, as described below.

\begin{figure}[h!]
\centering
\includegraphics[width=1.0\textwidth]{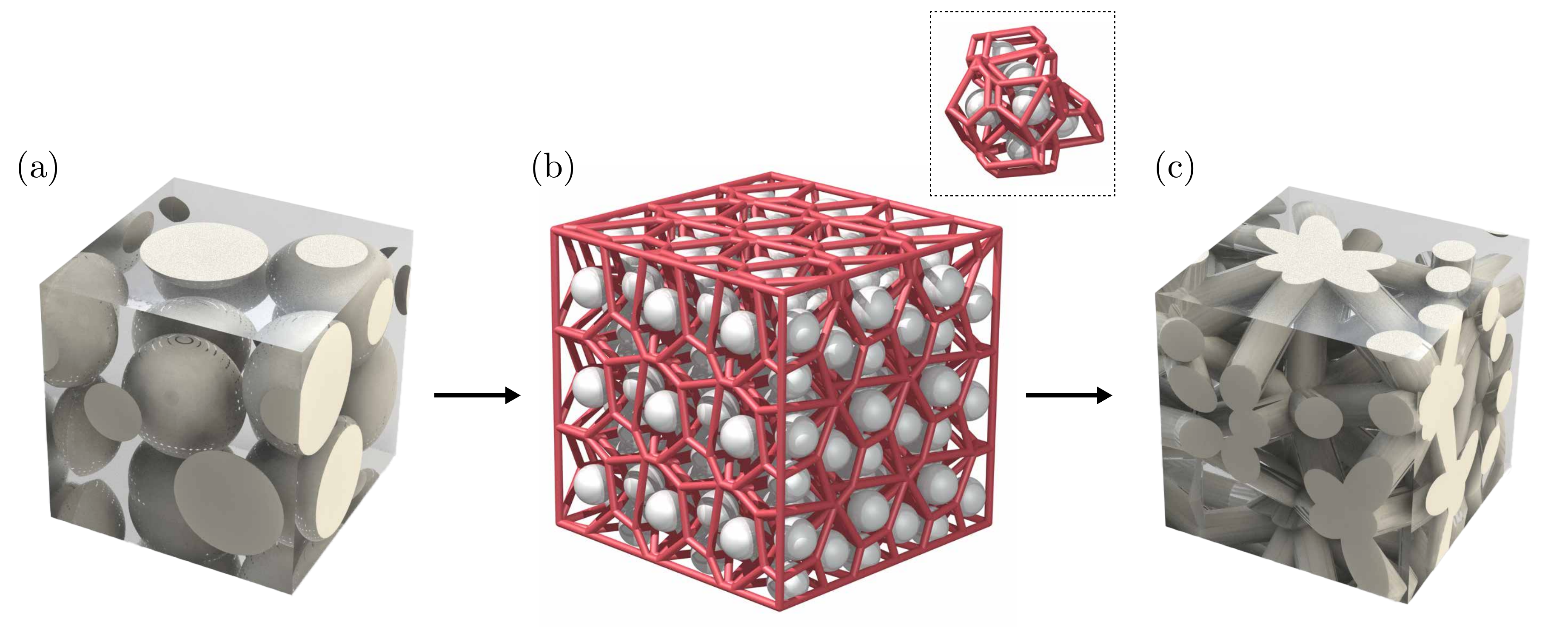} 
\vspace{-0.3in}
\caption{Constructing disordered representative volume elements for the micromechanical analysis: (a) RVE with dispersed hard domains, (b) Voronoi-tessellated 3 by 3 by 3 unit-cells (inset: Voronoi-tessellated unit-cell, where the center points of non-overlapping hard particles are used for tessellation), and (c) RVE with continuous hard domains. Here, v$_{\mathrm{hard}}=39.3 \%$ and $N=7$}
\label{fig:RVE}
\end{figure}

The center of the first particle is randomly placed inside a unit-cube. Then, the centers of the remaining particles are added to the unit cube. When placing the centers of the remaining particles, we loop over the centers of the existing particles and their 26 periodic images in order to check if the random trial coordinate of the center of the newly inserted particle is acceptable; i.e., the acceptable coordinate in the unit cube must be where the newly inserted particle does not overlap with the existing particles or any of 26 periodic images. Once the centers of the $N$ particles are appropriately distributed inside the unit-cube without any overlapping, we create 26 periodic images of the $N$ particles by offsetting from the original set of the $N$ particles. Lastly, a RVE is constructed, for which the central unit-cube containing the original $N$ particles was taken from 3 by 3 by 3 unit-cells. As expected, this RVE has three-dimensional periodicity without any overlapping of the particles. Figure \ref{fig:RVE}a shows an exemplar RVE with dispersed hard domains (hard volume fraction: 39.3\% and $N=7$) constructed using this simple algorithm. For the volume fractions considered here, this simple algorithm was acceptable with monodisperse hard particles. However, for higher volume fractions of hard components, more complicated algorithms for the mono- or poly-dispersed $N$ particles must be employed (e.g., \cite{torquato2002random, anoukou2018random, tarantino2019random}), which is beyond the scope of this work.
% Figure3

\subsubsection{Continuous hard domain}
To construct RVEs with continuous hard domains, we used Voronoi tessellation, which has been widely used to address the geometric features of diverse multi-phase or porous material systems such as cellular materials, foams, polycrystalline metals and granular materials (\cite{slotterback2008correlation}). Specifically, we used a standard three-dimensional Voronoi tessellation algorithm implemented in the open source code, Voro++ (\cite{rycroft2009voro++}) for RVEs with randomly connected, continuous hard domains.

First, as illustrated in Figure \ref{fig:RVE}b, we perform the Voronoi tessellation of $27N$ points of the 3 by 3 by 3 unit-cells. The 3 by 3 by 3 unit-cells were constructed using the 26 periodic images of the central unit-cell, where the $N$ points are taken to be the center points of the non-overlapping hard particles in the RVE with dispersed hard domains, as presented in Section \ref{section311}. Once the super-cell with the $27N$ center points (Voronoi points) is tessellated via a standard Voronoi algorithm, we obtain detailed information on neighbors (adjacent polyhedral cells sharing the tessellated planes) for each of the tessellated polyhedral sub-domains in the super-cell; in the inset of Figure \ref{fig:RVE}b, the Voronoi-tessellated central unit-cell is displayed as a visual aid. We then perform rod-connecting (with a constant radius; here, the radius is such that it meets the desired volume fraction of the hard domains) the center point and its neighboring points for all of the $N$ center points in the central unit-cube. When rod-connecting the center and the corresponding neighboring points, a cutoff radius for each of the polyhedral cells was taken to be 90\% of the maximum distance from the center point. Lastly, a RVE is constructed for which the central unit-cube was taken from the rod-connected structure. Figure \ref{fig:RVE}c shows an exemplar continuous RVE for a specified volume fraction (hard volume fraction: 39.3\% and $N=7$; more detailed information about this particular RVE example with seven random spatial points is provided in the Appendix \ref{appendixC}). As shown in the RVE, it has randomly connected local microstructures throughout the tessellated polyhedral network in and out of the central unit cube. Moreover, within the RVE, local microstructures are highly disordered while under periodic boundary conditions in all directions. The RVE is macroscopically periodic yet locally disordered.

\subsection{Micromechanical modeling setup}
After constructing the RVEs for the micromechanical analysis of the TPU materials, we conducted numerical simulations of RVEs subjected to appropriate boundary conditions. The boundary value problems for the RVEs were solved using finite elements within Abaqus/Standard. Toward this end, we numerically implemented the finite deformation constitutive models for both hard and soft constituent materials for use in the finite element solver. In the Appendix \ref{appendixA}, the nonlinear constitutive modeling framework and its computational implementation procedure are presented together with material parameters for the hard and soft constituent materials. In summary, a hyperelastic-viscoplastic constitutive model consisting of two micro-rheological mechanisms of a time-dependent hyperelastic-inelastic representation and a hyperelastic network representation was used for the hard constituent while a simple hyperelastic representation was used for the soft constituent. Furthermore, the material parameters for both components are extrapolated from the experimental data for the three TPU materials (TPU$_{\mathrm{A}}$, TPU$_{\mathrm{B}}$ and TPU$_{\mathrm{C}}$) with varying volume fractions of the hard and soft components.

\subsubsection{Periodic boundary condition}
Periodic boundary conditions were imposed on the RVEs with the dispersed or continuous hard domains constructed in Section \ref{section31}. Furthermore, to compute the macroscopic, average responses of the RVEs under periodic boundary conditions, we used the “fictitious node” virtual work method presented in \cite{danielsson2002three} and \cite{danielsson2007micromechanics}, which was developed using the principle of virtual work.

First, a three-dimensional periodic boundary condition for a pair of material points (A and B at the periodic pair surfaces) is imposed on the RVEs by,
\begin{equation}
\mathbf{u}_{A} - \mathbf{u}_{B} = \left( \bar{\mathbf{F}} - \mathbf{I}\right)\left(\mathbf{X}_{A} - \mathbf{X}_{B} \right),
\label{eq:pbc}
\end{equation}
where $\mathbf{u}$ is the local displacement, $\bar{\mathbf{F}}$ is the prescribed macroscopic deformation gradient imposed on the RVE, $\mathbf{X}$ is the material vector in the undeformed reference body, and $\mathbf{I}$ is the second-order identity tensor. Equation (\ref{eq:pbc}) gives the kinematic constraints for pairs of the material points on the periodic boundary surfaces. The overall, averaged stress response in the RVEs is then computed using the principle of virtual work such that an internal virtual work due to admissible variation in the displacement field balances the external virtual work; i.e., 
\begin{equation}
\delta W_{\mathrm{int}} = \delta W_{\mathrm{ext}}.
\end{equation}
The internal virtual work can be expressed in terms of the first Piola (engineering) stress and the deformation gradient (or displacement gradient), 
\begin{equation}
\delta W_{\mathrm{int}} = \int_{V_{0}} \mathbf{T}_{\mathrm{R}}(\mathbf{X}): \delta \mathbf{F}(\mathbf{X}) dV = V_{0} \bar{\mathbf{T}}_{\mathrm{R}}: \delta \bar{\mathbf{F}} = V_{0} \bar{\mathbf{T}}_{\mathrm{R}}: \delta \bar{\mathbf{H}},
\label{eq:Work_int}
\end{equation}
where $\mathbf{T}_{\mathrm{R}}$ is the local first Piola stress, $V_{0}$ is the volume at the “undeformed” reference configuration and $\mathbf{F} \equiv \frac{\partial \bm{\upvarphi}}{\partial \mathbf{X}}$ is the local deformation gradient that maps the material vector $\mathbf{X}$ at the reference to the spatial vector and $\mathbf{x}=\bm{\upvarphi}(\mathbf{X},t)$ at the deformed configuration, where $\bm{\upvarphi}$ is the motion. Furthermore, $\bar{\mathbf{T}}_{\mathrm{R}}$ is the macroscopic first Piola stress and $\bar{\mathbf{H}}=\bar{\mathbf{F}}-\mathbf{I}$ is the macroscopic displacement gradient. The external virtual work can be written as
\begin{equation}
\delta W_{\mathrm{ext}} = \int_{S_{0}} \mathbf{t}_{0}(\mathbf{X}) \cdot \delta \mathbf{u}(\mathbf{X}) dS = \int_{S_{0}} \mathbf{T}_{\mathrm{R}}(\mathbf{X}) \mathbf{n}_{0} \cdot \delta \mathbf{u}(\mathbf{X}) dS,
\label{eq:Work_ext}
\end{equation}
where $\mathbf{t}_{0} \equiv \mathbf{T}_{\mathrm{R}} \mathbf{n}_{0}$ is the surface traction, $\delta \mathbf{u}$ is the virtual displacement in the reference configuration and $\mathbf{n}_{0}$ is the outward unit vector normal to the surface. Meanwhile, the external virtual work can be computed by
\begin{equation}
\delta W_{\mathrm{ext}} = \mathbf{\Phi} : \delta \bm{\eta},
\label{eq:Work_general}
\end{equation}
where the components of $\bm{\eta}$ are the generalized degrees of freedom or the displacement components of the fictitious nodes (column vectors of $\bm{\eta}$) and where the column vectors of $\mathbf{\Phi}$ are the work conjugate generalized forces corresponding to the generalized degrees of freedom. Here, the admissible variation of the generalized degrees of freedom is identical to the variation of the macroscopic displacement gradient; i.e., $\delta \bm{\eta} = \delta \bar{\mathbf{H}}$. Hence, by using Equation (\ref{eq:Work_int}), (\ref{eq:Work_ext}) and (\ref{eq:Work_general}), the macroscopic “averaged” Piola stress in the RVE under a periodic boundary condition can be simply computed by
\begin{equation}
\bar{\mathbf{T}}_{\mathrm{R}} = \frac{1}{V_{0}}\int_{V_{0}} \mathbf{T}_{\mathrm{R}} (\mathbf{X}) dV = \frac{1}{V_{0}} \mathbf{\Phi},
\end{equation}
Moreover, the macroscopic Cauchy (true) stress is related to the macroscopic Piola stress by
\begin{equation}
\bar{\mathbf{T}} = \bar{J}^{-1} \bar{\mathbf{T}}_{\mathrm{R}} \bar{\mathbf{F}}^{\top}
\label{eq:macroCauchystress},
\end{equation}
where $\bar{J} = \mathrm{det} \bar{\mathbf{F}}$ is the macroscopic volume change in the RVE.

\subsubsection{Macroscopic loading condition}
In this work, the RVEs constructed in Section \ref{section31} are subjected to periodic boundary conditions as well as macroscopic uniaxial compression; i.e., 
\begin{equation}
\bar{\mathbf{F}} = \lambda_{1} \mathbf{e}_{1} \otimes \mathbf{e}_{1} + \lambda_{2} \mathbf{e}_{2} \otimes \mathbf{e}_{2} + \bar{\lambda} \mathbf{e}_{3} \otimes \mathbf{e}_{3}
\end{equation}
where $\bar{\lambda}$ is the prescribed stretch in the loading direction ($\mathbf{e}_{3}$) and $\lambda_{1}$ and $\lambda_{2}$ are the unconstrained lateral stretches determined in the boundary value problems for the RVEs.

\subsubsection{Identification of the number of random Voronoi points for representative volume elements}
\begin{figure}[h!]
\vspace{0.2in}
\centering
\includegraphics[width=1.0\textwidth]{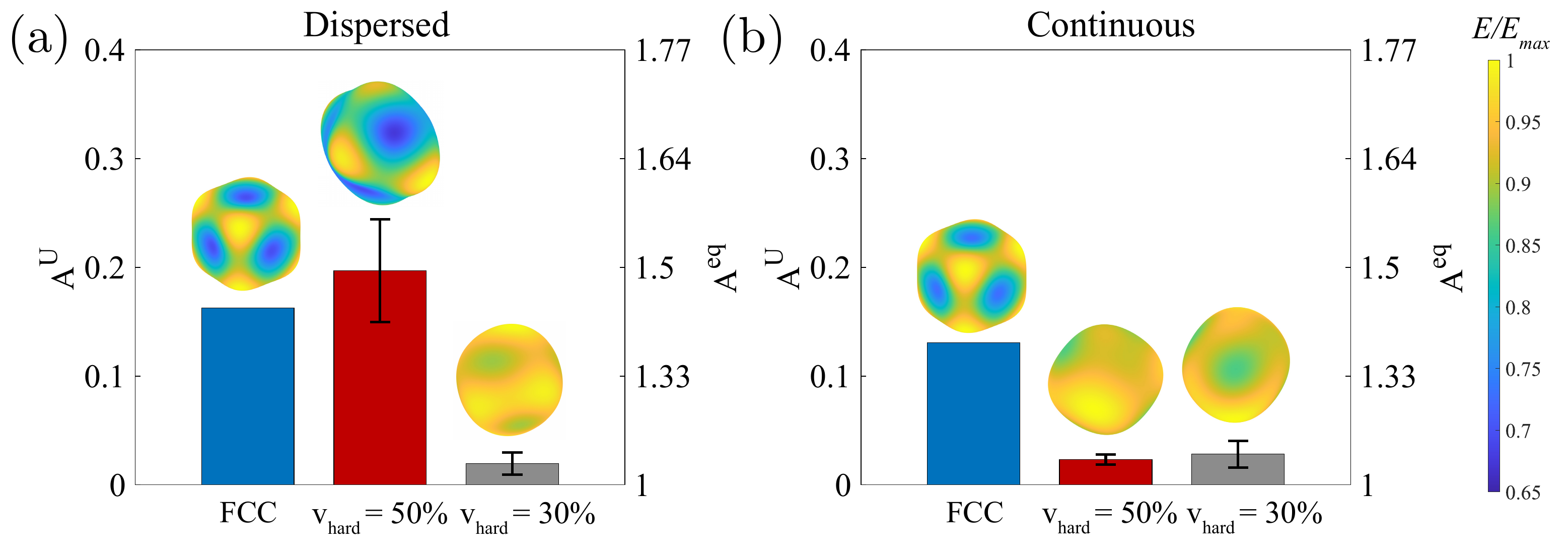}
\vspace{-0.3in}
\caption{Anisotropy in disordered RVEs with (a) dispersed hard domains and (b) continuous hard domains used in micromechanical analysis. Here, A$^{\mathrm{U}}$ is the universal anisotropy index and A$^{\mathrm{eq}}$ is the equivalent Zener index (\cite{ranganathan2008universal,nye1985physical}). Anisotropy indices are displayed together with sphere-like anisotropy maps that represent “relative” stiffness throughout crystallographic orientations. Additionally, anisotropy maps for face-centered-cubic (FCC) materials with dispersed and continuous hard domains are presented for comparisons with the disordered RVEs.}
\label{fig:anisotropy}
\end{figure}

In the following section \ref{section4}, we present micromechanical analysis results for the Voronoi tessellation-based disordered RVEs with the finite number of Voronoi points under periodic boundary conditions. Ideally, the number of Voronoi points should be large enough to achieve well homogenized, isotropic behaviors of the RVEs. However, as the number of Voronoi points increases, packing the non-overlapping hard spheres (v$_{\mathrm{hard}}=$26.9 \%, 39.3 \% and 52.2 \%) considered in this study becomes much more complicated. Furthermore, it is extremely difficult to properly mesh the RVEs with a large number of hard spheres. Hence, choosing the appropriate number of Voronoi points is crucial for the “efficient” micromechanical analysis. In this study, the number of Voronoi points was chosen to be $N=7$. As shown in Figure \ref{fig:anisotropy} on universal anisotropy index (\cite{ranganathan2008universal}), the RVEs with $N=7$ exhibited reasonably low anisotropy. The RVEs with the continuous hard domains exhibited “near” complete isotropy at both v$_{\mathrm{hard}}=30 \%$ and v$_{\mathrm{hard}}=50 \%$, but the RVE with the dispersed hard domains at the higher volume fraction of the hard component (v$_{\mathrm{hard}}=$ 50\%) was found to exhibit significant anisotropy with a large variation in the anisotropy index. If the number of Voronoi points increases ($N>7$), anisotropy in this RVE would diminish. However, we have not increased the number of Voronoi points but simply chosen $N=7$ for more efficient micromechanical analysis. 

\section{Experiments vs. micromechanical modeling results}
\label{section4}
Here, we present micromechanical modeling results for the RVEs with highly disordered microstructures presented in Section \ref{section3}. 

% Figure5
\begin{figure}[t!]
\centering
% \hspace{-0.3in}
\includegraphics[width=0.65\textwidth]{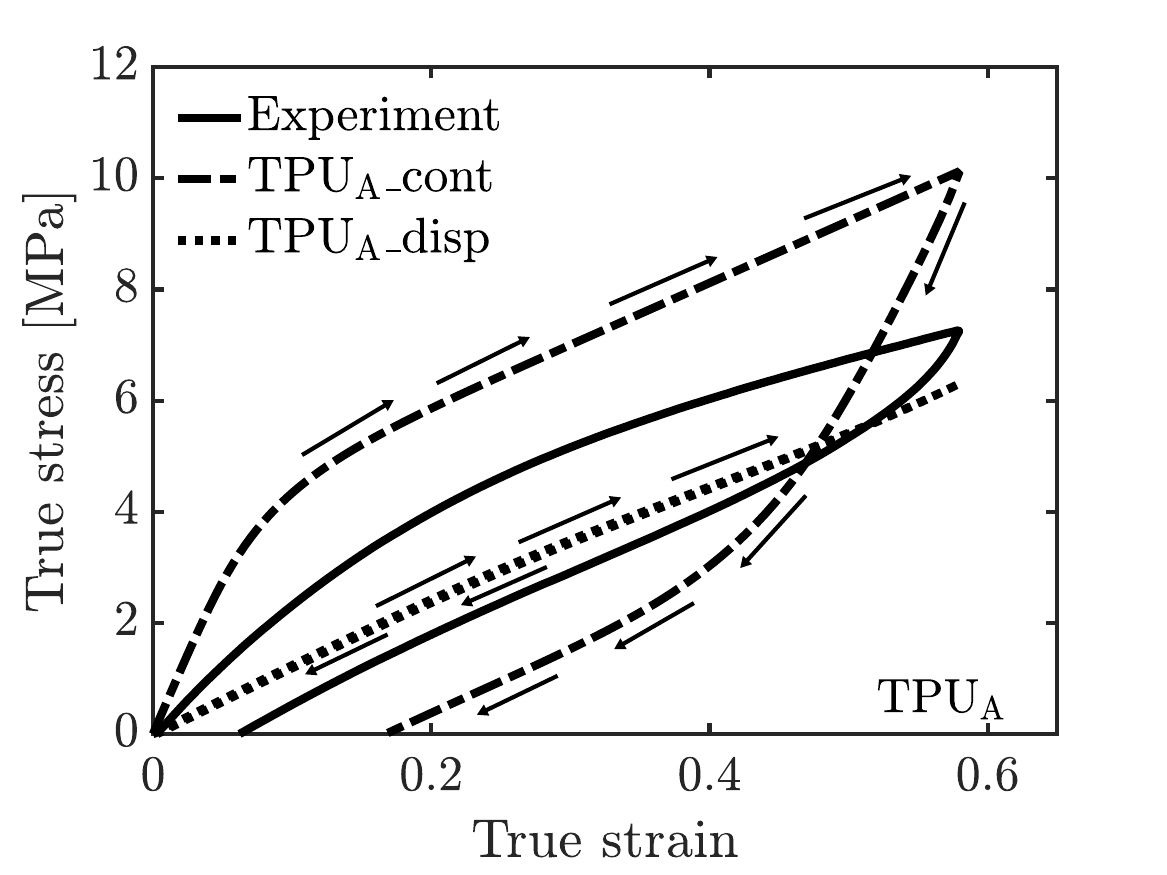} 
\vspace{-0.1in}
\caption{Experiment vs. micromechanical modeling results for TPU$_{\mathrm{A}}$ (v$_{\mathrm{hard}}=26.9 \%$) under cyclic loading and unloading conditions at a strain rate of 0.01 $\mathrm{s}^{-1}$. Geometric information for this TPU$_{\mathrm{A}}$ is given in the Appendix \ref{appendixC}.}
\label{fig:tpuA_load_unload}
\end{figure}

% TPU$_{\mathrm{A}}$\_cont 

Figure \ref{fig:tpuA_load_unload} presents the micromechanical modeling results for TPU$_{\mathrm{A}}$\_cont (v$_{\mathrm{hard}}=26.9 \%$) together with the corresponding experimental data under uniaxial compressive loading and unloading condition. The macroscopic, average stress response in the RVE with continuous hard domains was found to be greater than that for the RVE with dispersed hard domains. The applied load was transferred more efficiently throughout the hard ligament network, which resulted in greater stress response in the RVE with continuous hard domains from an early stage of loading. In addition to the much stiffer initial elastic response, the stress rollover due to yield was more apparent in TPU$_{\mathrm{A}}$\_cont, compared to its dispersed-particle counterpart, TPU$_{\mathrm{A}}$\_disp. More interestingly, the stress-strain curve from the experimental data for TPU$_{\mathrm{A}}$ was found to be located between those in the micromechanical modeling results for TPU$_{\mathrm{A}}$\_disp and TPU$_{\mathrm{A}}$\_cont. It should also be noted that there was no hysteresis in TPU$_{\mathrm{A}}$\_disp while a significant amount of energy was dissipated in TPU$_{\mathrm{A}}$\_cont during the cycle. In addition to the energy dissipation, elastic shape recovery at the end of unloading in the experiment was also observed between those from TPU$_{\mathrm{A}}$\_cont and TPU$_{\mathrm{A}}$\_disp. These micromechanical modeling results for the two different morphologies imply that, with a relatively low hard content, both dispersed and continuous microstructures can be present in the TPU material. More specifically, the Young’s modulus in experiment ($E_{\mathrm{exp}}$) was found to be $E_{\mathrm{exp}} \sim 0.67E_{\mathrm{disp}} + 0.33E_{\mathrm{cont}}$, where $E_{\mathrm{disp}}$ and and $E_{\mathrm{cont}}$ are the Young’s moduli from the micromechanical models for TPU$_{\mathrm{A}}$\_disp and TPU$_{\mathrm{A}}$\_cont, respectively. Therefore, it may be argued that this TPU material comprises $\sim 67 \,\%$ of the dispersed microstructures and $\sim 33 \,\%$ of the continuous microstructures.

\begin{figure}[t!]
\centering
% \hspace{-0.3in}
\includegraphics[width=0.65\textwidth]{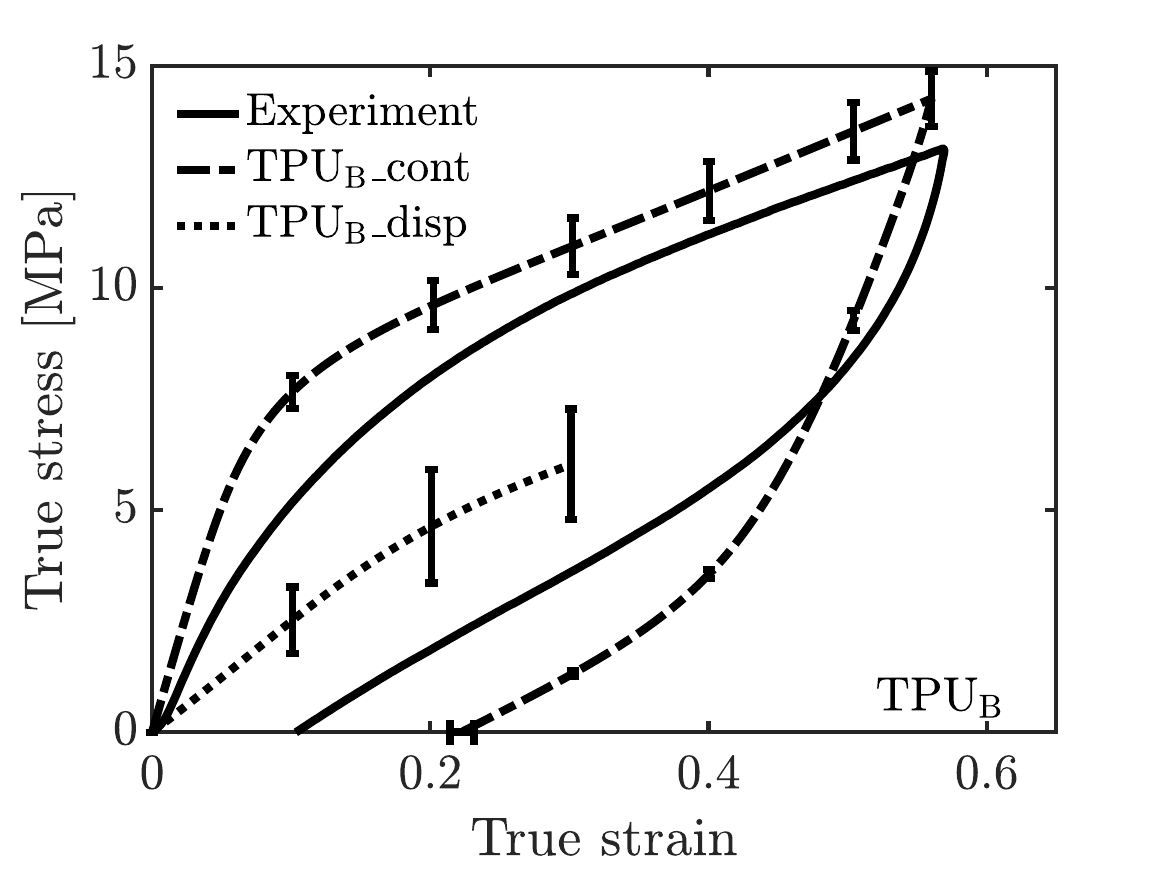}
\vspace{-0.1in}
\caption{Experiment vs. micromechanical modeling results for TPU$_{\mathrm{B}}$ (v$_{\mathrm{hard}}=39.3 \%$) under cyclic loading and unloading conditions at a strain rate of 0.01 $\mathrm{s}^{-1}$.}
\label{fig:tpuB_load_unload}
\end{figure}

Comparisons of the experimental data and micromechanical modeling results are further presented in Figure \ref{fig:tpuB_load_unload} for TPU$_{\mathrm{B}}$ (v$_{\mathrm{hard}}=39.3 \%$).\footnote{We constructed five independent RVEs for the dispersed-particle morphology and its continuous counterpart and conducted numerical simulations of the RVEs for the statistical treatment of the micromechanical modeling results for TPU$_{\mathrm{B}}$.} The experimentally observed stress-strain curve for TPU$_{\mathrm{B}}$ was closer to the macroscopic stress-strain response with continuous hard domains from small to large strains.\footnote{The numerical simulations for TPU$_{\mathrm{B}}$\_disp were not able to converge beyond a macroscopically imposed stain of 0.3 due to severe deformation in the surrounding soft matrices. 
} This implies that TPU$_{\mathrm{B}}$ with a relatively higher volume fraction of hard components is more likely to possess continuous hard microstructures. More specifically, the Young’s modulus in experiment ($E_{\mathrm{exp}}$) was found to be $E_{\mathrm{exp}} \sim 0.43E_{\mathrm{disp}} + 0.57E_{\mathrm{cont}}$, where $E_{\mathrm{disp}}$ and and $E_{\mathrm{cont}}$ are the Young’s moduli from the micromechanical models for TPU$_{\mathrm{B}}$\_disp and TPU$_{\mathrm{B}}$\_cont, respectively. Therefore, it may be argued that this TPU material comprises $\sim 43 \,\%$ of the dispersed microstructures and $\sim 57 \,\%$ of the continuous microstructures.

% Figure6

\begin{figure}[t!]
\centering
% \hspace{-0.3in}
\includegraphics[width=0.65\textwidth]{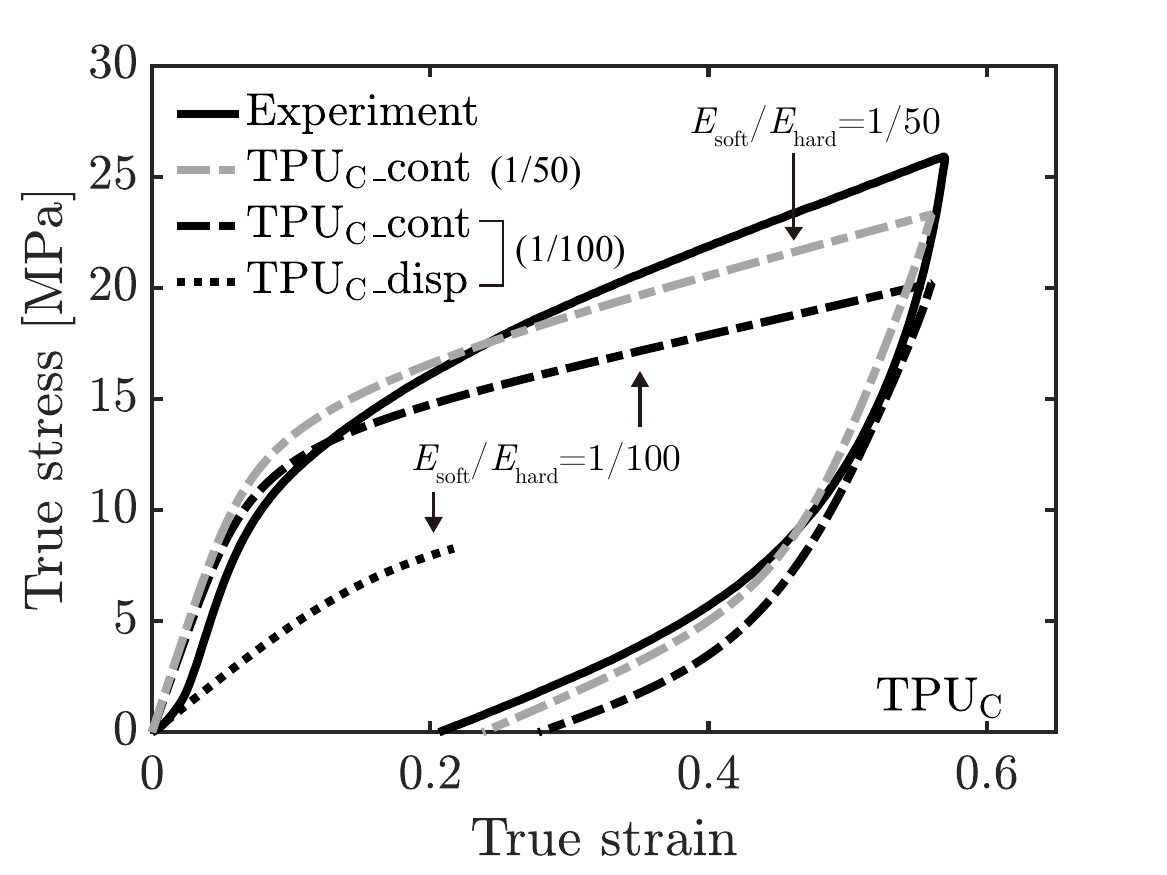} 
\vspace{-0.1in}
\caption{Experiment vs. micromechanical modeling results for TPU$_{\mathrm{C}}$ (v$_{\mathrm{hard}}=52.2 \%$) under cyclic loading and unloading conditions at a strain rate of 0.01 $\mathrm{s}^{-1}$. Geometric information for this TPU$_{\mathrm{C}}$ is given in the Appendix \ref{appendixC}.}
\label{fig:tpuC_load_unload}
\end{figure}

The resilient yet dissipative cyclic behavior is further examined for TPUc with the highest volume fraction of hard components in Figure \ref{fig:tpuC_load_unload}. Here, the micromechanical modeling results are displayed for both dispersed and continuous morphologies.\footnote{The numerical simulations for TPU$_{\mathrm{C}}$\_disp were not able to converge beyond a macroscopically imposed stain of 0.2 due to severe deformation in the surrounding soft matrices} Though the micromechanical model prediction with continuous hard domains nicely matched the experimental data at the early stage of loading, there was a clear discrepancy between the experimental data and model predictions beyond the yield. Uncertainty in the constitutive behaviors in both hard and soft constituents can also lead to this discrepancy (see the Appendix \ref{appendixA} for details about the constitutive behaviors for hard and soft components). This is further supported by the micromechanical modeling results with two different initial stiffness ratios ($E_{\mathrm{soft}}/E_{\mathrm{hard}}$) of 1/100 (black dashed line) and 1/50 (gray dashed line), where the micromechanical model prediction with a greater elastic modulus in the soft constituent ($E_{\mathrm{soft}}/E_{\mathrm{hard}} = 1/50$) better matched the experimental data in terms of flow stresses at increasing strains and overall features in shape recovery and hysteresis in TPU$_{\mathrm{C}}$. 

% Figure7

% Figure8
\begin{figure}[t!]
\centering
\includegraphics[width=0.9\textwidth]{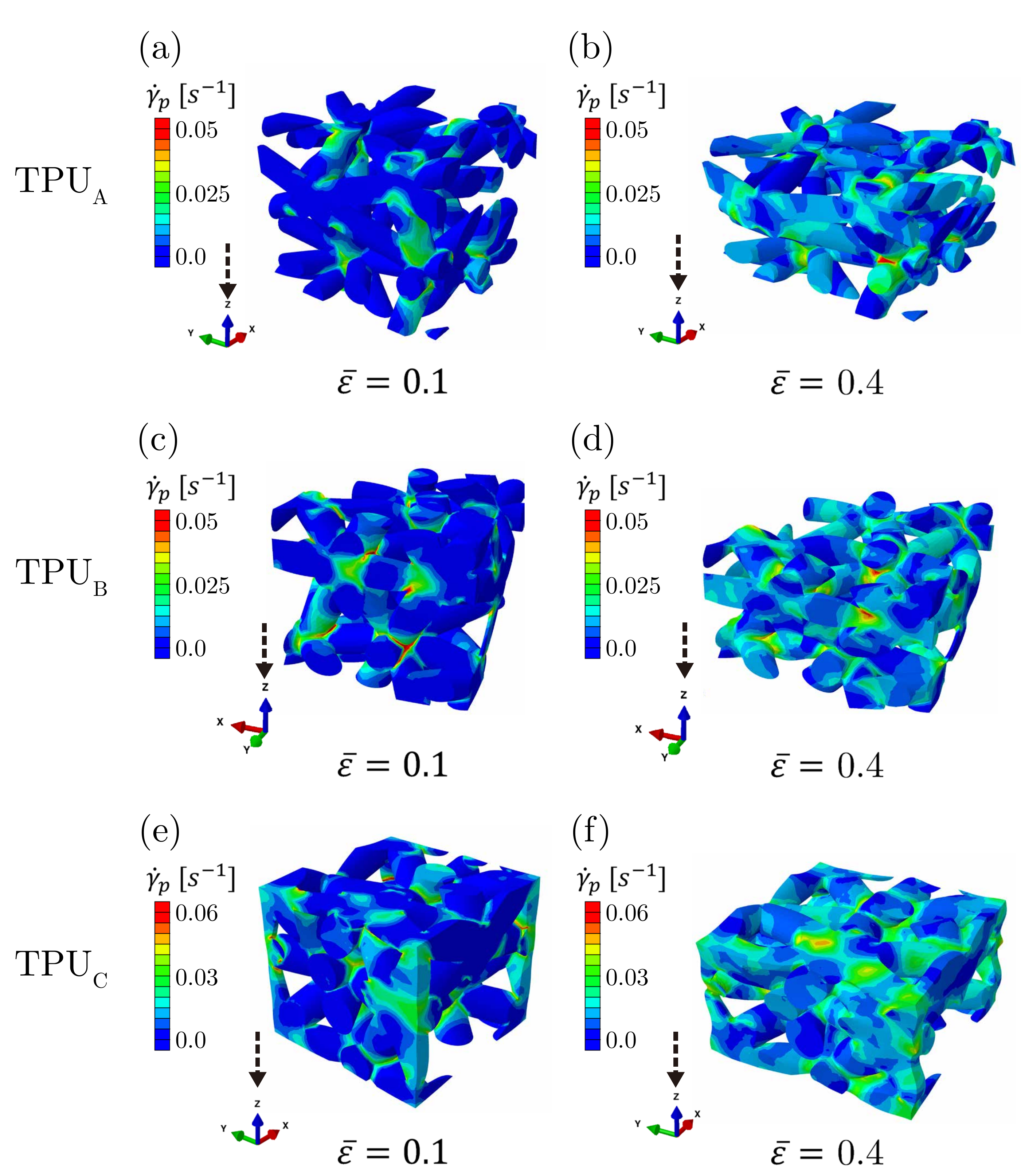} 
\caption{Contours of plastic strain rates throughout the hard ligament networks in micromechanical models. TPU$_{\mathrm{A}}$ at (a) $\bar{\varepsilon}$ = 0.1, (b) $\bar{\varepsilon}$ = 0.4, TPU$_{\mathrm{B}}$ at (c) $\bar{\varepsilon}$ = 0.1, (d) $\bar{\varepsilon}$ = 0.4 and TPU$_{\mathrm{C}}$ at (e) $\bar{\varepsilon}$=0.1, (f) $\bar{\varepsilon}$=0.4. Here, only the hard domains are displayed for a visual aid.}
\label{fig:contour_gd}
\end{figure}

% Figure9
\begin{figure}[t!]
\centering
\includegraphics[width=1.0\textwidth]{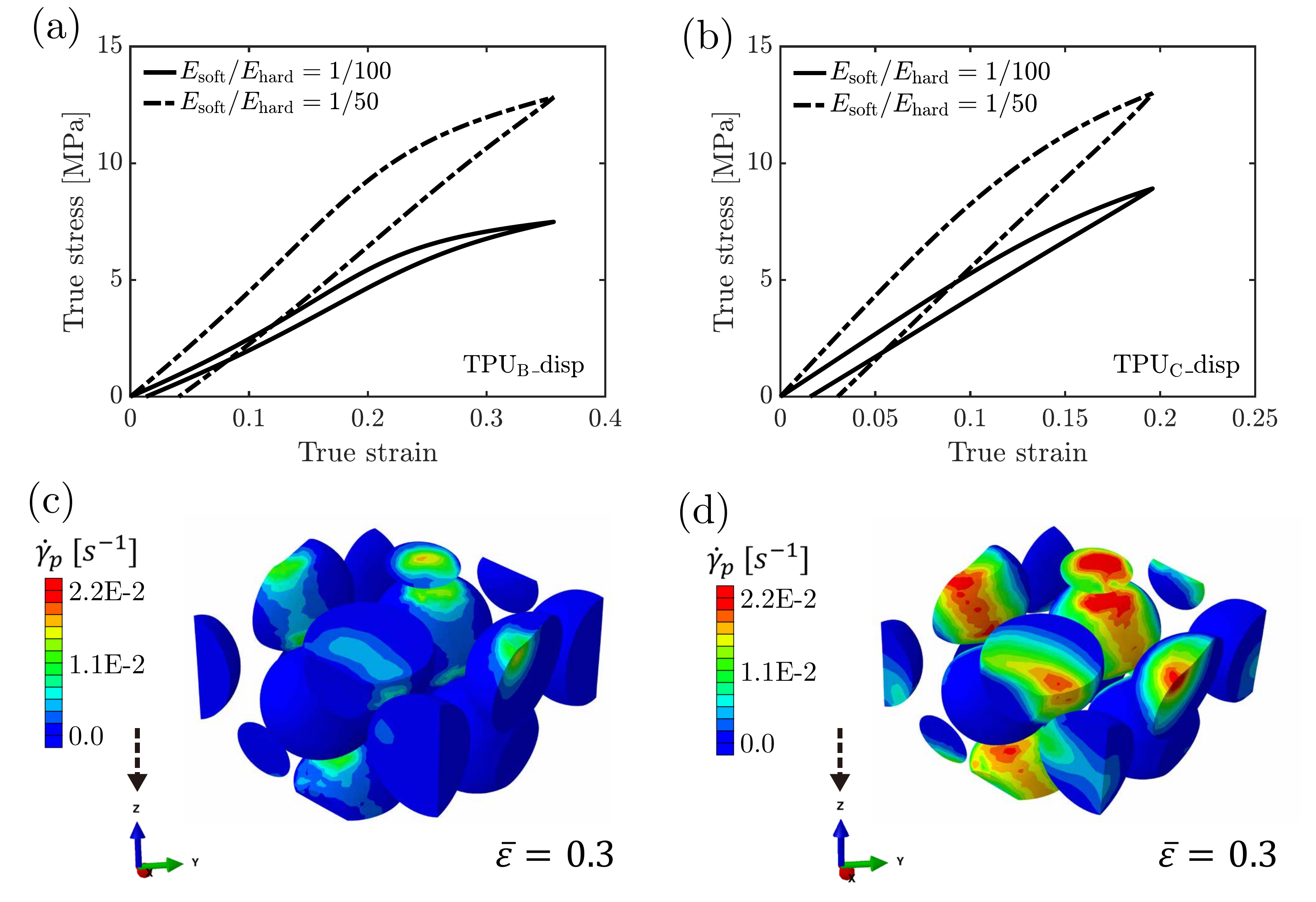}
\vspace{-0.2in}
\caption{Micromechanical modeling results for TPU$_{\mathrm{B}}$\_disp and TPU$_{\mathrm{C}}$\_disp with two different stiffness ratios ($E_{\mathrm{soft}}/E_{\mathrm{hard}}$) of 1/50 and 1/100. Stress-strain curves of (a) TPU$_{\mathrm{B}}$\_disp and (b) TPU$_{\mathrm{C}}$\_disp under cyclic loading and unloading conditions. Contours of plastic strain rates throughout the hard particles in the micromechanical models for TPU$_{\mathrm{B}}$\_disp with two stiffness ratios of (c) $E_{\mathrm{soft}}/E_{\mathrm{hard}}$=1/100 at $\bar{\varepsilon}$ = 0.3 and (d) $E_{\mathrm{soft}}/E_{\mathrm{hard}}$=1/50 at $\bar{\varepsilon}$ = 0.3. }
\label{fig:disp_load_unload}
\end{figure}

% -------------------------------------------
Further detailed micromechanical modeling results are presented in Figure \ref{fig:contour_gd}. Figure \ref{fig:contour_gd}a, Figure \ref{fig:contour_gd}c and Figure \ref{fig:contour_gd}e depict the contours of plastic strain rates throughout the continuous hard domains in TPU$_{\mathrm{A}}$\_cont, TPU$_{\mathrm{B}}$\_cont and TPU$_{\mathrm{C}}$\_cont at an imposed macroscopic strain of 0.1. Beyond the initial elastic regime, the plastic flow began to develop throughout the hard ligament networks, which culminates in the stress-rollover observed apparently in the macroscopic, average stress-strain curves in the RVEs with continuous hard domains (Figure \ref{fig:tpuA_load_unload}, \ref{fig:tpuB_load_unload} and \ref{fig:tpuC_load_unload}). The plastic flow developed further throughout the continuous hard domains as the imposed macroscopic strain was increased to 0.4, as displayed in Figure \ref{fig:contour_gd}b, Figure \ref{fig:contour_gd}d and Figure \ref{fig:contour_gd}f. Throughout all deformation processes at small to large strains, the hard ligament network plays a crucial role in the inelastic features observed in all of the TPU materials. By contrast, the stress-rollover due to yield was not apparent in the micromechanical modeling results for the RVEs with dispersed hard domains (TPU$_{\mathrm{A}}$\_disp, TPU$_{\mathrm{B}}$\_disp and TPU$_{\mathrm{C}}$\_disp), as shown in Figure \ref{fig:tpuA_load_unload}, \ref{fig:tpuB_load_unload} and \ref{fig:tpuC_load_unload}. Furthermore, as shown in the numerically simulated stress-strain curve displayed in Figure \ref{fig:disp_load_unload}a, no significant energy was dissipated in TPU$_{\mathrm{B}}$\_disp during a loading and unloading cycle. As shown in Figure \ref{fig:disp_load_unload}c on the contours of the plastic strain rates in the RVE with dispersed hard particles, a plastic flow with a much smaller magnitude was localized throughout two adjacent hard particles. Even at an imposed macroscopic strain of 0.3, the dispersed hard domains underwent no significant plastic deformation and most of the imposed deformation was carried by the hyperelastic soft domains. To examine the role of the soft domains in inelastic deformation features in the RVEs with dispersed hard particles further, we present the numerical simulation results for TPU$_{\mathrm{B}}$\_disp with two different initial stiffness ratios of soft to hard components of 1/100 (solid line) and 1/50 (dashed line), as shown in Figure \ref{fig:disp_load_unload}a, for which the material parameters of the hard constituent were fixed. Highly hysteretic behavior with more energy dissipation was available in TPU$_{\mathrm{B}}$\_disp with a greater elastic modulus in the soft constituent ($E_{\mathrm{soft}}/E_{\mathrm{hard}} = 1/50$). This is further evidenced by the plastic strain rate contour in the RVE displayed in Figure \ref{fig:disp_load_unload}c and Figure \ref{fig:disp_load_unload}d. The hard particles exhibited stronger plastic flow since more of the imposed macroscopic deformation is accommodated by the hard particles in the RVE with $E_{\mathrm{soft}}/E_{\mathrm{hard}} = 1/50$. Additionally, the micromechanical modeling results are presented for TPU$_{\mathrm{C}}$\_disp with two different initial stiffness ratios of soft to hard components of 1/100 (solid line) and 1/50 (dashed line) in Figure \ref{fig:disp_load_unload}b. The overall elastic-inelastic features are similar to those in TPU$_{\mathrm{B}}$\_disp; the hysteretic stress-strain behavior accompanied by more energy dissipation was better captured in the RVE with the stiffer soft constituent ($E_{\mathrm{soft}}/E_{\mathrm{hard}} = 1/50$). However, stress-rollover due to yield in the hard particles was found to occur at an earlier stage of loading ($\bar\varepsilon\sim$ 0.13) in TPU$_{\mathrm{C}}$\_disp than in TPU$_{\mathrm{B}}$\_disp ($\bar\varepsilon\sim$ 0.22), further supporting the critical role of the soft domains in determining the inelastic properties of dispersed-particle morphologies.

% % Figure11
% \begin{figure}[b!]
% \centering
% \includegraphics[width=1.0\textwidth]{fig/contour2}
% \caption{Contours of evolving network elasticity modulus and shear strength throughout the hard ligament network in TPU$_{\mathrm{C}}$\_cont in consecutive cycles of loading, unloading and reloading. (a) Network elasticity modulus and (b) plastic shear strength. }
% \label{fig:contour2}
% \end{figure}
% Figure10
\begin{figure}[b!]
\centering
\includegraphics[width=1.0\textwidth]{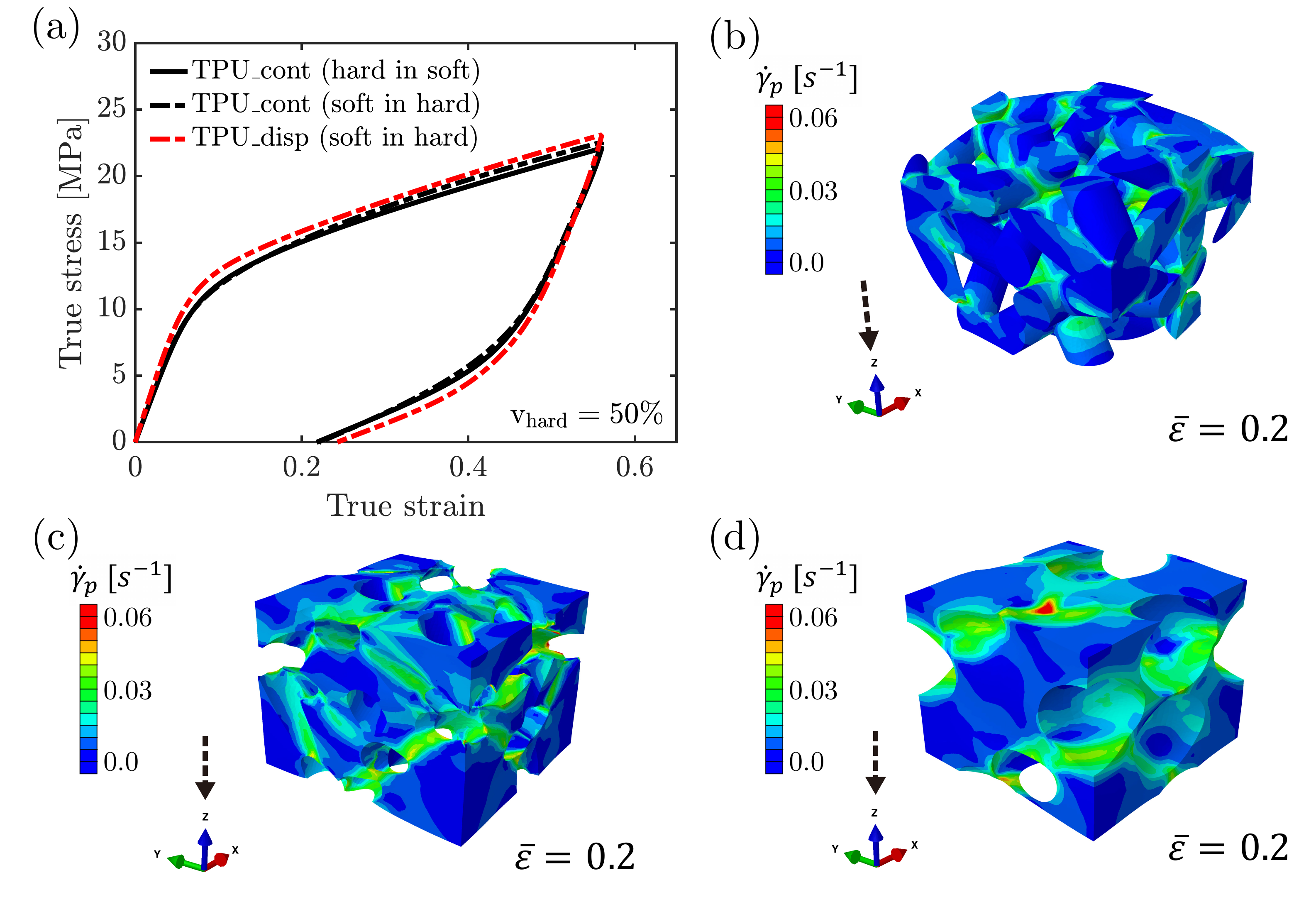} 
\vspace{-0.2in}
\caption{Micromechanical modeling results for RVEs with different compositions. (a) Macroscopic stress-strain curves in RVEs with rod-connected microstructures (\textit{hard-in-soft} and \textit{soft-in-hard}) and a RVE with dispersed particles (\textit{soft-in-hard}). Contours of plastic strain rates in (b) RVE with rod-connected microstructures: \textit{hard-in-soft}, (c) RVE with rod-connected microstructures: \textit{soft-in-hard} (d) RVE with dispersed particles: \textit{soft-in-hard} at a macroscopic strain of $\bar{\varepsilon}=0.2$. Here, v$_{\mathrm{hard}}=50 \%$. Geometric information for these TPU\_disp and TPU\_cont (v$_{\mathrm{hard}}=50 \%$) is given in the Appendix \ref{appendixC}.}
\label{fig:soft_in_hard}
\end{figure}

As shown in the micromechanical modeling results on TPU$_{\mathrm{A}}$, TPU$_{\mathrm{B}}$ and TPU$_{\mathrm{C}}$ in Figures \ref{fig:tpuA_load_unload}, \ref{fig:tpuB_load_unload}, \ref{fig:tpuC_load_unload} and \ref{fig:contour_gd}, connectivity throughout the subdomains plays a critical role in determining elastic and inelastic characteristics in these materials. The role of connectivity is further explored in Figure \ref{fig:soft_in_hard} on the simulation results for RVEs with different compositions. First, we conducted numerical simulations for RVEs with rod-connected microstructures (v$_{\mathrm{hard}}=50 \%$): \textit{hard-in-soft} (hard components in soft matrices) and \textit{soft-in-hard} (soft components in hard matrices). Here, it should be noted that RVEs with \textit{hard-in-soft} composition have been used for all of the micromechanical modeling results for TPU$_{\mathrm{A}}$, TPU$_{\mathrm{B}}$ and TPU$_{\mathrm{C}}$ presented in Figure \ref{fig:tpuA_load_unload}, \ref{fig:tpuB_load_unload}, \ref{fig:tpuC_load_unload}, \ref{fig:contour_gd} and \ref{fig:disp_load_unload}. As shown in Figure \ref{fig:soft_in_hard}a, interestingly, the macroscopic stress-strain curve (black dashed line) from the \textit{soft-in-hard} composition was found to be very close to that (black solid line) in the RVE with the \textit{hard-in-soft} composition. These simulation results are very reasonable since this RVE has “co-continuous” morphology, where both domains are continuous and interpenetrating with no significant isolated domains. Furthermore, the overall features in the macroscopic stress-strain curve (red dashed line) from a dispersed-particle RVE with the \textit{soft-in-hard} composition are also very similar to those from the RVEs with rod-connected microstructures. These results clearly support that connectivity throughout the hard domains is of critical importance in the highly resilient yet dissipative mechanical features in these TPU materials with the high fraction of the hard thermoplastic components. This is further demonstrated in Figures \ref{fig:soft_in_hard}b, \ref{fig:soft_in_hard}c and \ref{fig:soft_in_hard}d on the contours of plastic strain rates throughout the continuous, hard ligament networks in the RVEs with rod-connected microstructures \textit{(hard-in-soft} and \textit{soft-in-hard}) and the dispersed-particle RVE (\textit{soft-in-hard}). The plastic flows developed well throughout the hard ligament networks in all three RVEs. Moreover, the magnitudes in the plastic strain rates were found to be comparable in all three RVEs, which resulted in very similar behaviors in these RVEs.

% Figure11
\begin{figure}[b]
\centering
\includegraphics[width=1.0\textwidth]{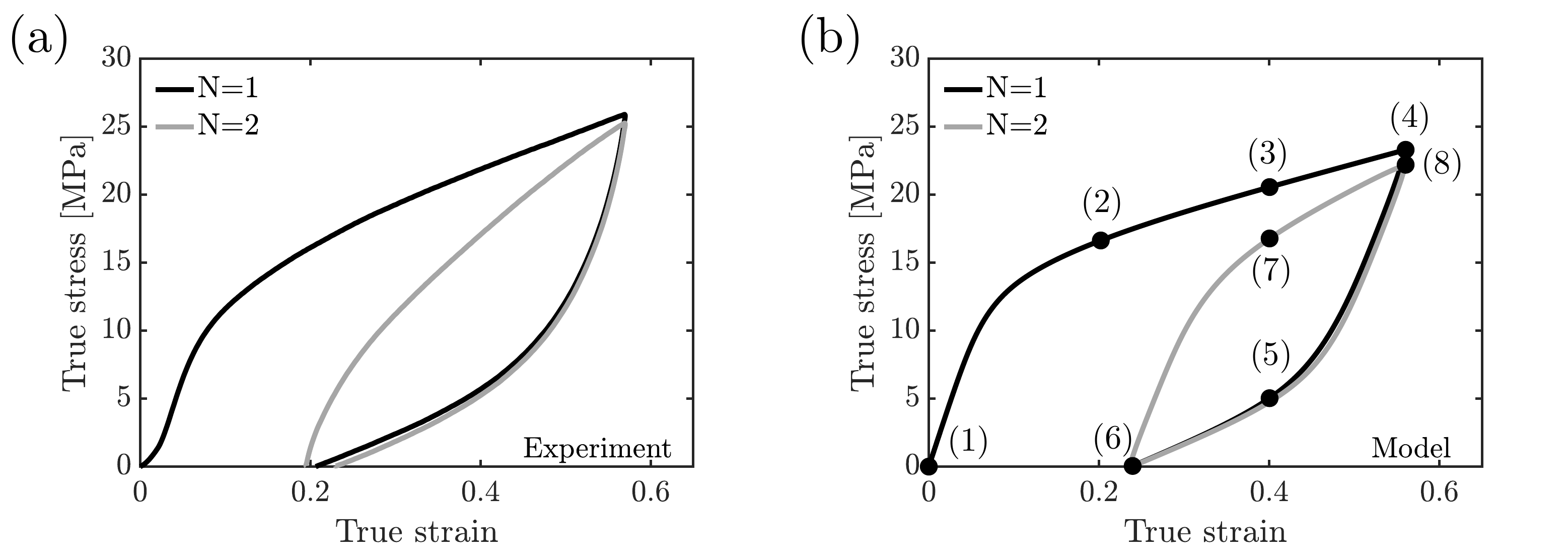} 
\vspace{-0.3in}
\caption{Experiment vs. micromechanical modeling results for TPU$_{\mathrm{C}}$ during multiple, consecutive cycles of loading, unloading, reloading and unloading at a strain rate of 0.01 $\mathrm{s}^{-1}$. (a) Experimental data, (b) macroscopic, average stress-strain curves in the micromechanical model. }
\label{fig:cycle}
\end{figure}
\begin{figure}[h!]
\centering
\includegraphics[width=0.9\textwidth]{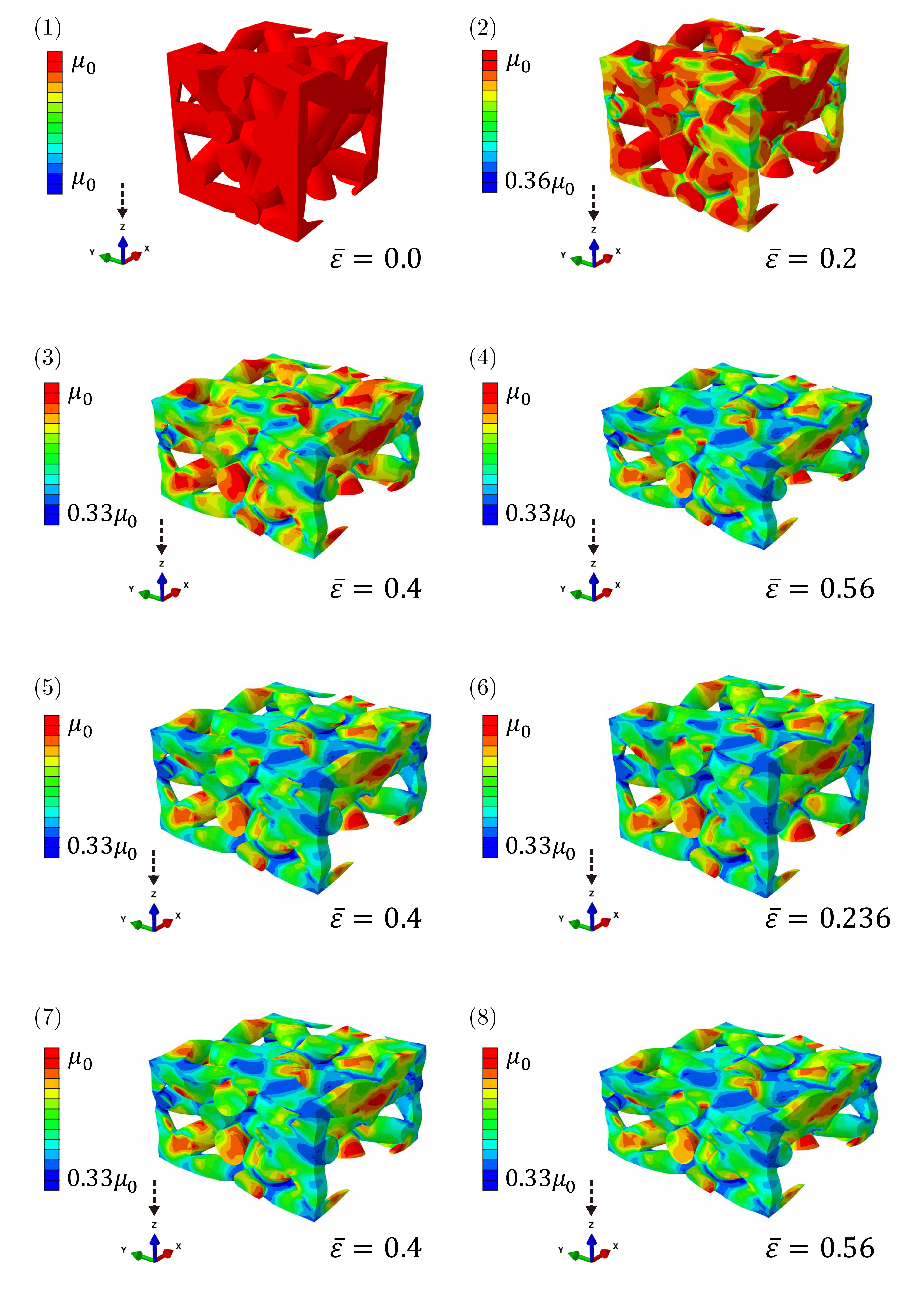}
%\vspace{-0.3in}
\caption{Contours of the evolving network elasticity modulus throughout the hard ligament network in TPU$_{\mathrm{C}}$\_cont during consecutive cycles of loading, unloading and reloading.}
\label{fig:contour_muN}
\end{figure}
\begin{figure}[h!]
\centering
\includegraphics[width=0.9\textwidth]{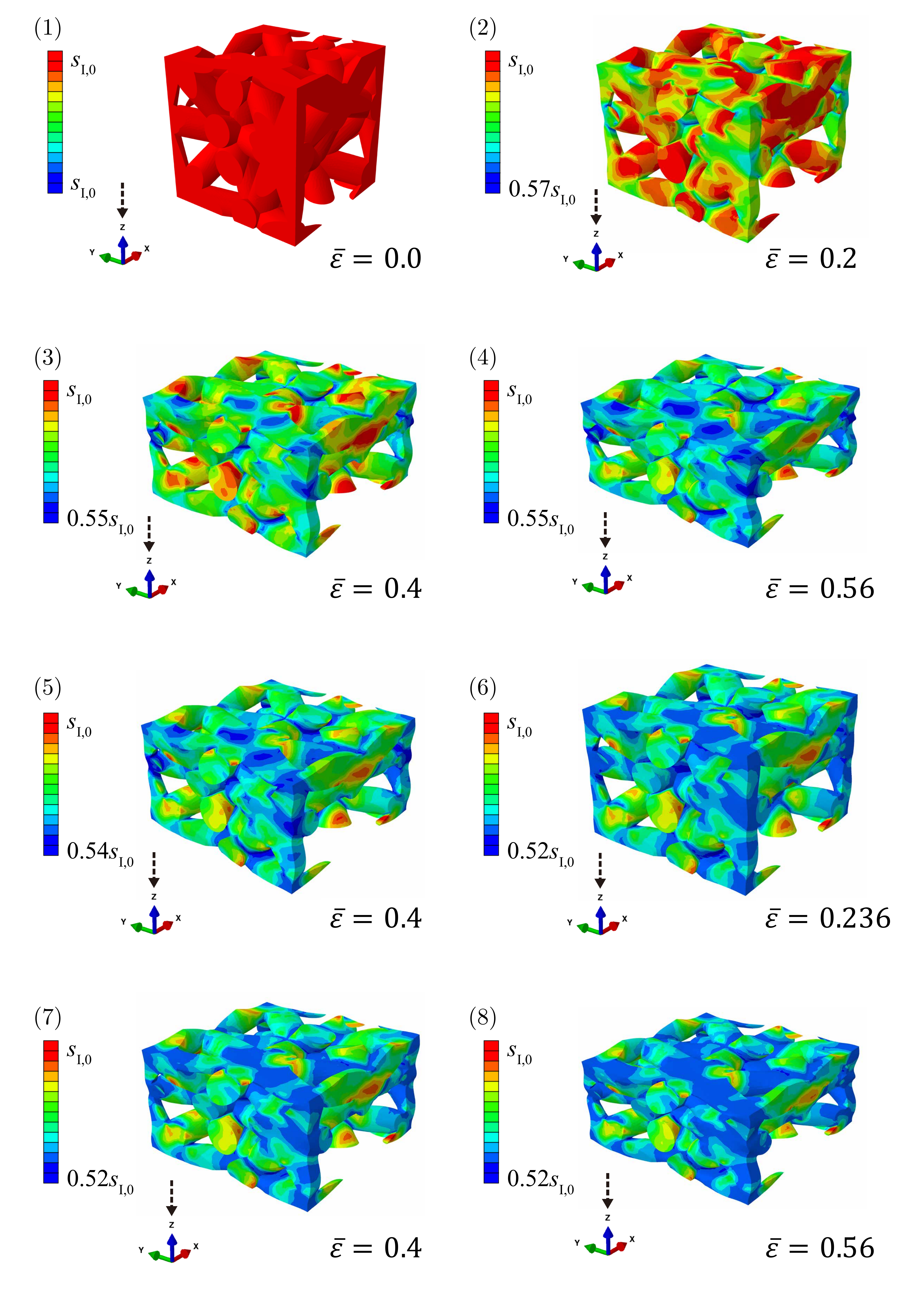}
\vspace{-0.3in}
\caption{Contours of evolving plastic shear strength throughout the hard ligament network in TPU$_{\mathrm{C}}$\_cont during consecutive cycles of loading, unloading and reloading.}
\label{fig:contour_sI}
\end{figure}

The micromechanical modeling results are presented together with the corresponding experimental data on multiple consecutive loading, unloading and reloading cycles for the TPU material with the highest volume fraction of hard components in Figure \ref{fig:cycle}. Highly hysteretic mechanical features were observed in both cycles with remarkably dissipated energy as well as significant elastic shape recovery in both experiments and micromechanical models. The numerically simulated stress-strain curves (Figure \ref{fig:cycle}b) display good agreement with the experimental data (Figure \ref{fig:cycle}a) in the first and second cycles. More interestingly, stretch-induced softening, also known as the Mullins’ effect, was clearly manifested in the second cycle and was nicely captured by the micromechanical model. In the second cycle, the overall stress response lessened in both experiments and numerical simulations. The Mullins’ effect is mainly attributed to the microstructural breakdown throughout the hard domains (\cite{rinaldi2011microstructure, rinaldi2011tunable}) and is more evidenced in the contours of the network elasticity modulus and plastic shear strength presented in Figure \ref{fig:contour_muN} and Figure \ref{fig:contour_sI}, respectively. During the first loading up to a maximum imposed strain of 0.56, the network elastic modulus ($\mu_\mathrm{N}$; see Appendix \ref{appendixA} for details about constitutive models for hard and soft constituents) gradually decreases throughout the hard domains. It should be noted that the network elastic modulus decreased more significantly in the junctions of the hard ligament network. Furthermore, the elastic softening was found to be more pronounced in the hard ligaments under tensile bending modes. More interestingly, in the micromechanical modeling results, the network elastic modulus throughout the hard ligament network did not evolve during the unloading stage in the first cycle and during the reloading and unloading stages in the second cycle. In addition to the softened network elastic modulus during the cyclic deformation conditions, the plastic shear strength that resists the plastic flow decreased throughout the hard ligament network, as shown in Figure \ref{fig:contour_sI}. The decrease in the plastic shear strength ($s_\mathrm{I}$; see the Appendix \ref{appendixA} for the details on constitutive models for hard and soft constituents) is also responsible for the significantly softened stress responses in the second cycle. Note that while the network elastic modulus did not evolve, the plastic shear strength decreased during the unloading, reloading and unloading stages, by which the stress level at the maximum imposed strain of 0.56 in the second cycle (at (8) in Figure \ref{fig:cycle}b) was slightly less than that in the first cycle (at (4) in Figure \ref{fig:cycle}b), which is observed in the experimental data in Figure \ref{fig:cycle}a.  

%\clearpage
% Figure12

\clearpage
% Figure13

\begin{figure}[t!]
\centering
\includegraphics[width=1.0\textwidth]{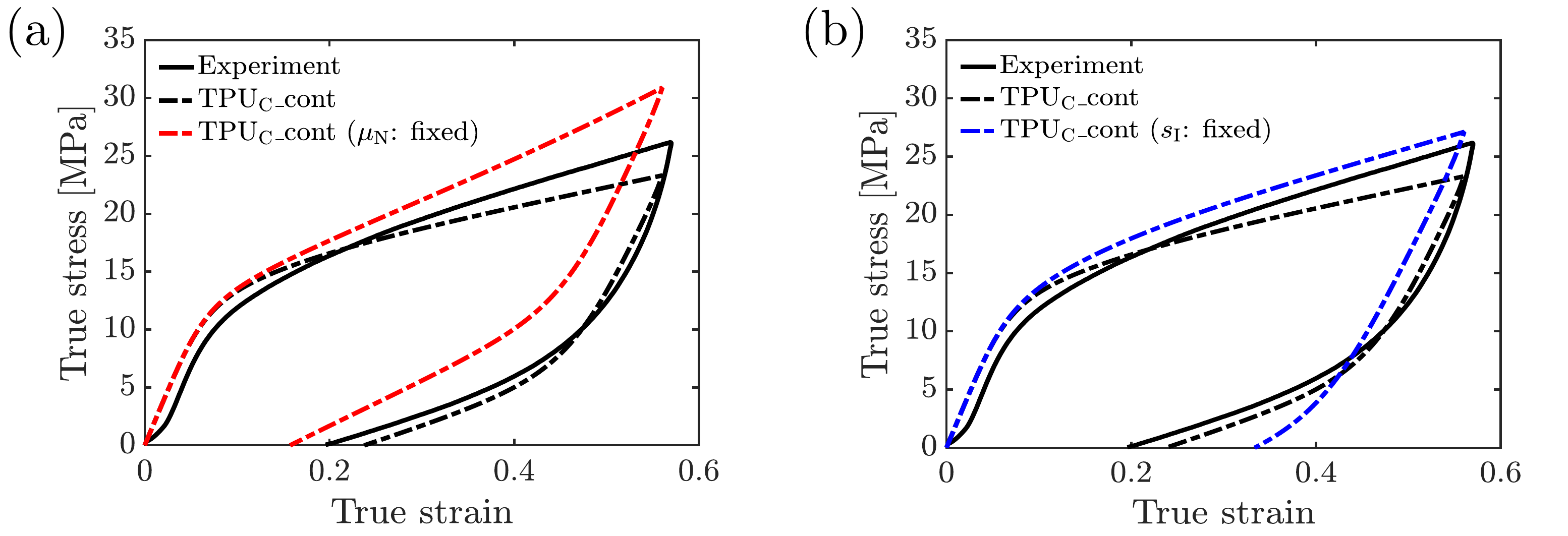} 
\vspace{-0.2in}
\caption{Micromechanical modeling results for TPU$_{\mathrm{C}}$ with and without softening mechanisms. Macroscopic stress-strain curves in (a) experiment vs. TPU$_{\mathrm{C}}$\_cont with and without evolution in the network elastic modulus, $\mu_\mathrm{N}$ and (b) experiment vs. TPU$_{\mathrm{C}}$\_cont with and without evolution in the plastic shear strength, $s_\mathrm{I}$ in the hard domains.}
\label{fig:mullins}
\end{figure}

The deformation-induced softening mechanism is further demonstrated in Figure \ref{fig:mullins}. Simulation results with and without evolution in the network elastic modulus in the hard domains due to the microstructural breakdown are presented together with experimental data in Figure \ref{fig:mullins}a. With no evolution in the network elastic modulus, the macroscopic stress response in the micromechanical model was found to be much greater than the experimental data, especially beyond the initial yield. By contrast, the residual strain at the end of unloading in the micromechanical model with no network softening decreased significantly, compared to that with network softening. The simulation results are further presented with and without evolution in the plastic shear strength in the hard domains in Figure \ref{fig:mullins}b. The overall stress response in the micromechanical model with no plastic shear strength evolution better matched the experimental data during “loading”. However, the unloading behavior was poorly predicted in the micromechanical model with no plastic shear strength evolution. These simulation results more clearly support that both elastic and inelastic softening mechanisms, here represented by evolutions in $\mu_{\mathrm{N}}$ and $s_{\mathrm{I}}$, play crucial roles in overall stress responses as well as in resilience and dissipation upon loading and unloading conditions.
Overall, the micromechanical models provided microstructural details for the elastic-plastic softening mechanisms widely reported in macroscopic cyclic deformation experiments in  many multi-phase thermoplastic elastomeric materials (\cite{rinaldi2011microstructure, cho2013constitutive, boyce2001deformation, diani2009review, lee2023polyurethane}). 

\section{Discussion and conclusion}
\label{section5}
Highly resilient and dissipative large strain behavior is the hallmark of thermoplastic elastomers. It is mainly attributed to the presence of phase-separated morphologies of hard and soft domains. Hence, the geometric features of hard and soft domains and their impact on resilience and dissipation have long been attractive for engineering the mechanics of two-phase thermoplastic elastomeric materials subjected to diverse deformation conditions. In this work, we have presented a micromechanical modeling framework to offer a microscopic understanding of the underlying geometric and morphological mechanisms for a broad range of elastic and inelastic features experimentally reported in representative thermoplastic polyurethane (TPU) materials possessing a wide range of volume fractions of hard and soft contents. To this end, new representative volume elements (RVE) that mimic the more realistic disordered microstructures often found in TPU materials were proposed. Using Voronoi tessellation for spatially random points, two distinct disordered RVEs were constructed, where the hard domains are either dispersed or connected with no isolated domains. More specifically, using Voronoi tessellation algorithms under three-dimensional periodic boundary conditions, detailed information about the neighboring points for each of the spatially random points was available, by which highly disordered yet randomly connected microstructures were newly designed. 

We then conducted micromechanical analyses of three different TPU materials using the RVEs with statistically random microstructures for specified volume fractions of hard and soft domains and then directly compared the micromechanical modeling results with the corresponding experimental stress-strain data in diverse loading scenarios reported in our previous work (\cite{cho2017deformation}). As presented in Section \ref{section4}, the micromechanical modeling results are in good agreement with the experimental data for the TPU materials. At relatively higher volume fractions of hard components, RVEs with continuous hard domains exhibited better predictive capabilities. However, at a lower hard volume fraction, the macroscopically observed stress-strain data were found to lie between the micromechanical modeling results with the dispersed-particle morphologies and their continuous counterparts. More interestingly, connectivity throughout the hard domains plays a key role in all elastic and inelastic features, including the initial stiffness, plastic yield, flow stresses, energy dissipation and shape recovery upon cyclic loading and unloading conditions. The macroscopically imposed deformation was found to be efficiently accommodated throughout the hard ligament network in RVEs with randomly connected microstructures from the early stage of loading, which resulted in relatively stiff initial elastic responses followed by more apparent yield behavior. Overall, the RVEs with the hard ligament network were found to be capable of better capturing the main features in the highly resilient and dissipative mechanical behaviors as well as in the stretch-induced elastic-inelastic softening experimentally found in these TPU materials. 

It should be noted that the overall macroscopic stress-strain behaviors observed in our micromechanical models for the TPU materials are very similar to those in thermoplastic vulcanizate (TPV) materials in Boyce and coworkers (\cite{boyce2001deformation}, \cite{boyce2001micromechanisms} and \cite{boyce2001micromechanics}). The TPV materials are not copolymers but phase-separated polymers, where “soft” ethylene propylene diene monomer (EPDM) elastomeric particles are dispersed or occluded in a “hard” continuous thermoplastic polypropylene domain. The TPV microstructures are quite different from the TPU microstructures but these two thermoplastic elastomers exhibit very similar mechanical behaviors especially upon cyclic loading and unloading conditions. Interestingly, the dispersed-particle morphology with \textit{soft-in-hard} composition (Figure \ref{fig:soft_in_hard}), where the hard thermoplastic domain was continuous, is very similar to that in the TPV materials. Furthermore, the average, macroscopic stress-strain curves and the plastic strain rate contours were very similar in all micromechanical modeling results for the RVEs with rod-connected microstructures (\textit{hard-in-soft} and \textit{soft-in-hard}) and the dispersed-particle RVE (\textit{soft-in-hard}) at v$_{\mathrm{hard}}=50 \%$, as presented in Figure \ref{fig:soft_in_hard}a, b, c and d. All of these micromechanical modeling results in both TPU and TPV clearly imply that connectivity throughout the hard thermoplastic domains plays a critical role in determining the highly resilient yet dissipative large strain behavior in these two-phase elasto-plastomeric materials. Furthermore, the co-continuous, interpenetrating microstructures often found in the TPU materials studied in this work are recently utilized to enhance the load transfer, energy dissipation and shape recovery capabilities in multi-component, heterogeneous materials architected on crystal lattices with long-range order (\cite{cho2016engineering,wang2011co,lee2012periodic,li2018enhanced}).

Lastly, though the proposed RVEs with either continuous or dispersed hard domains are highly disordered, they are intrinsically anisotropic due to the possible local symmetry of the underlying microstructures of particles or ligament networks. Hence, the choice of loading direction can influence the average responses predicted in the micromechanical models, as briefly illustrated in Figure \ref{fig:anisotropy} on anisotropy maps of the RVEs used in our micromechanical analysis. The anisotropy maps for the disordered microstructures in terms of the numbers of the random spatial points as well as the volume fractions of hard and soft components should be further analyzed in the next steps. Furthermore, more interestingly, the average number of connectivities (or neighboring points to each of the random spatial points) in the RVEs with $N=7$ presented in Section \ref{section4} was found to be $\sim$ 12, very close to that in the close-packed face-centered-cubic (FCC) system. The number of connectivities and the corresponding impact on the anisotropy maps as well as the large strain elastic-inelastic features in these disordered microstructures can also be investigated in future work. More specifically, the proposed disordered microstructures with hard ligament networks can be utilized for designing near-complete isotropic materials with the “finite” number of random spatial points able to reconcile the thermodynamically conflicting nature of resilience and energy dissipation under extreme loading conditions. Furthermore, high-resolution multi-material three-dimensional printing technology can be employed in order to realize highly disordered, heterogeneous materials with superb resilience and dissipation as well as with near-complete isotropy. In turn, the micromechanical modeling framework and the new RVEs presented in this work should enable the designs and assessments of other disordered, cellular or multi-phase materials at both microscopic and macroscopic length-scales. 

%see the macroscopic, average stress-strain curves with statistical error bars for TPU$_{\mathrm{B}}$ presented in Figure \ref{fig:tpuB_C_load}a and Figure \ref{fig:tpuB_C_load_unload}b

\section*{Acknowledgement}
\textcolor{white}{.}
The authors gratefully acknowledge financial support from BASF. H.C. also acknowledges financial support provided by National Research Foundation of Korea (2021R1A4A103278312).

\renewcommand*\appendixpagename{Appendix}
\renewcommand*\appendixtocname{Appendix}
\begin{appendices}
\numberwithin{equation}{section}
\numberwithin{figure}{section}
\section{Constitutive behavior of hard and soft constituents and their computational implementation}
\label{appendixA}
\subsection{Constitutive equations}
Here, the constitutive behaviors of hard and soft components are outlined. Detailed information on the large strain constitutive models for TPU$_{\mathrm{A}}$, TPU$_{\mathrm{B}}$ and TPU$_{\mathrm{C}}$ comprising (1) a time-dependent elastic-inelastic micro-rheological mechanism (denoted I) and (2) an equilibrium hyperelastic network micro-rheological mechanism (denoted N) in our previous work, \cite{cho2017deformation}. The material parameters used in the constitutive model for the pure hard component were simply extrapolated from those in TPU$_{\mathrm{A}}$ (v$_{\mathrm{hard}}=26.9 \%$), TPU$_{\mathrm{B}}$ (v$_{\mathrm{hard}}=39.3 \%$) and TPU$_{\mathrm{C}}$ (v$_{\mathrm{hard}}=52.2 \%$), using simple power-law fitting procedures since there are no data available for pure hard components in these TPU materials. The constitutive equations and material parameters for the hard component are summarized, as follows.

%% Kinematics
A motion $\bm\upvarphi$ is defined as a one-to-one mapping $\mathbf{x}=\bm\upvarphi(\mathbf{X},t)$ with a material point $\mathbf{X}$ in a fixed undeformed reference and $\mathbf{x}$ in a deformed spatial configuration. The large deformation kinematics for elastic-inelasticity in the mechanisms I follows a multiplicative decomposition of a deformation gradient $\mathbf{F}\defeq\frac{\partial\bm\upvarphi}{\partial\mathbf{X}}$,
\begin{equation}
\mathbf{F}=\mathbf{F}^{e}_{\mathrm{I}}\mathbf{F}^{p}_{\mathrm{I}}=\mathbf{F}_{\mathrm{N}}
\end{equation}
where $\mathbf{F}^{e}_{\mathrm{I}}$ and $\mathbf{F}^{p}_{\mathrm{I}}$ are the elastic and inelastic deformation gradients for the time-dependent mechanism I. We then define the following basic kinematic tensors  for the hard component:
\begin{equation}
\begin{aligned}
& \mathbf{F}^{e}_{\mathrm{I}} = \mathbf{R}^{e}_{\mathrm{I}} \mathbf{U}^{e}_{\mathrm{I}}, && \text{polar decomposition of } \, \mathbf{F}^{e}_{\mathrm{I}} \, \text{into rotation and stretch;}  \\
& \bar{\mathbf{F}}_{\mathrm{N}} = J^{-1/3}\mathbf{F}_{\mathrm{N}} ,  && \text{isochoric part of } \, \mathbf{F}_{\mathrm{N}}  \; \text{where} \; J\defeq\det\mathbf{F}_{\mathrm{N}}; \\
%& \text{isochoric part of } \, \mathbf{C}^{e} && \Bar{\textbf{C}}^{e(\alpha)}=\bar{\mathbf{F}}^{{e(\alpha)}\top}\bar{\mathbf{F}}^{e(\alpha)} \\
& \bar{\mathbf{B}}_{\mathrm{N}} =\bar{\mathbf{F}}_{\mathrm{N}} \bar{\mathbf{F}}^{\top}_{\mathrm{N}}, && \text{isochoric left Cauchy-Green tensor.} \\
\end{aligned}
\end{equation}

\noindent The deformation rate is described by the velocity gradient $\mathbf{L}\defeq\mathrm{grad}\textbf{v}=\frac{\partial \textbf{v}}{\partial \textbf{x}}$ decomposed into elastic and inelastic parts,
\begin{equation}
\begin{aligned}
\mathbf{L} & =\dot{\mathbf{F}}\mathbf{F}^{-1} \\
& =\dot{\mathbf{F}}^{e}_{\mathrm{I}}\mathbf{F}^{e-1}_{\mathrm{I}}+\mathbf{F}^{e}_{\mathrm{I}}\dot{\mathbf{F}}^{p}_{\mathrm{I}}\mathbf{F}^{p-1}_{\mathrm{I}}\mathbf{F}^{e-1}_{\mathrm{I}} \\
& = \mathbf{L}^{e}_{\mathrm{I}}+\mathbf{F}^{e}_{\mathrm{I}}\mathbf{L}^{p}_{\mathrm{I}}\mathbf{F}^{e-1}_{\mathrm{I}}.
\end{aligned}
\end{equation}

\noindent The symmetric part of $\mathbf{L}^{p}_{\mathrm{I}}$ is the rate of inelastic stretching, $\mathbf{D}^{p}_{\mathrm{I}}$; and the skew part of $\mathbf{L}^{p}_{\mathrm{I}}$ is the inelastic spin, $\mathbf{W}^{p}_{\mathrm{I}}$; i.e., $\mathbf{L}^{p}_{\mathrm{I}}=\mathbf{D}^{p}_{\mathrm{I}}+\mathbf{W}^{p}_{\mathrm{I}}$. Furthermore, we make two important kinematical assumptions for inelastic flow; the flow is incompressible, i.e. $J^{p}_{\mathrm{I}}\defeq\det\mathbf{F}^{p}_{\mathrm{I}}=1$ and irrotational, i.e. $\mathbf{W}^{p}_{\mathrm{I}}=0$. Thus, the rate of inelastic deformation gradient is expressed by
\begin{equation}
\dot{\mathbf{F}}_{\mathrm{I}} = \mathbf{D}^{p}_{\mathrm{I}}\mathbf{F}^{p}_{\mathrm{I}}. 
\label{eq:Fpderivative}
\end{equation}

\noindent Then, the total stress tensor in the hard component is obtained by,
\begin{equation}
\mathbf{T}_{\mathrm{hard}} = \mathbf{T}_{\mathrm{I}} + \mathbf{T}_{\mathrm{N}},
\end{equation}
where $\mathbf{T}_{\mathrm{I}}$ and $\mathbf{T}_{\mathrm{N}}$ are the Cauchy stresses in the mechanisms I and N, respectively. The Cauchy stress in the time-dependent elastic-inelastic mechanism I is expressed by,
\begin{equation}
\begin{aligned}
\mathbf{T}_{\mathrm{I}} = \frac{1}{J} \mathbf{R}^{e}_{\mathrm{I}} \mathbf{M}^{e}_{\mathrm{I}} \mathbf{R}^{e \top}_{\mathrm{I}} \qquad \text{where} \quad \mathbf{M}^{e}_{\mathrm{I}} = 2\mu_{\mathrm{I}} (\ln \mathbf{U}^{e}_{\mathrm{I}})_{0} \, + \, K(\ln J)\mathbf{I}.
\end{aligned}
\end{equation}

\noindent We note that the bulk response is lumped into the mechanism I. Here, the rate of inelastic stretching $\mathbf{D}^{p}_{\mathrm{I}}$ is coaxial to the deviatoric Mandel stress, $(\mathbf{M}^{e}_{\mathrm{I}})_{0} = \mathbf{M}^{e}_{\mathrm{I}} - \frac{1}{3} (\mathrm{tr} \mathbf{M}^{e}_{\mathrm{I}}) \mathbf{I}$,
\begin{equation}
\mathbf{D}^{p}_{\mathrm{I}} =\dot{\gamma}^{p}_{\mathrm{I}}\mathbf{N}^{p}_{\mathrm{I}} \quad \text{with} \quad \mathbf{N}^{p}_{\mathrm{I}} = \frac{(\mathbf{M}^{e}_{\mathrm{I}})_{0}}{\|(\mathbf{M}^{e}_{\mathrm{I}})_{0}\|}.
\end{equation}

\noindent We then employed the thermally-activated viscoplasticity model prescribed by,
\begin{equation}
\begin{aligned}
\dot{\gamma}^{p}_{\mathrm{I}} = \dot{\gamma}_{\mathrm{I},0} \exp \left(- \frac{\Delta G_{\mathrm{I}}}{k_{B} \theta} \right) \sinh \left(\frac{\Delta G_{\mathrm{I}}}{k_{B} \theta} \frac{\tau_{\mathrm{I}}}{s_{\mathrm{I}}} \right) \quad \text{where} \quad \tau_{\mathrm{I}}= \frac{1}{\sqrt{2}}\|(\mathbf{M}^{e}_{\mathrm{I}})_{0}\| \\
\end{aligned}
\label{eq:gdflowrule}
\end{equation}
\noindent
with the pre-exponential factor $\Dot{\gamma}_{0,\mathrm{I}}$, the activation energy $\Delta{G}_{\mathrm{I}}$, Boltzmann's constant $k$, the absolute temperature $\theta$, and the shear strength $s_{\mathrm{I}}$. Furthermore, to capture the inelastic strain softening beyond yield induced by a reduction of the intermolecular resistance due to rearrangement of the molecular structures during inelastic flow in hard domains, we employed a simple saturation-type evolution rule for $s_{\mathrm{I}}$, 
\begin{equation}
\dot{s}_{\mathrm{I}} = h_{\mathrm{I}} \left(1-\frac{s_{\mathrm{I}}}{s_{\mathrm{I},ss}} \right)\dot{\gamma}^{p}_{\mathrm{I}}
\label{eq:sevolution}
\end{equation}
where $h_{\mathrm{I}}$ is the softening slope and the initial value of shear strength $s_{\mathrm{I},0}=17.6$ MPa.

In the meantime, the Cauchy stress in the time-independent hyperelastic mechanism N is given by,
\begin{equation}
\mathbf{T}_{\mathrm{N}} = \frac{\mu_{\mathrm{N}}}{3J} \frac{\lambda_{L}}{\bar{\lambda}_{\mathrm{N}}} \mathscr{L}^{-1} \left(\frac{\bar{\lambda}_{\mathrm{N}}}{\lambda_{L}}\right) (\bar{\mathbf{B}}_{\mathrm{N}})_{0} \qquad \text{where} \quad \bar{\lambda}_{\mathrm{N}} = \sqrt{\frac{\text{tr}\bar{\mathbf{B}}_{\mathrm{N}}}{3}},
\end{equation}
with the shear modulus $\mu_{\mathrm{N}}$ and the limiting chain extensibility $\lambda_{L}$. Also, $\mathscr{L}^{-1}$ is the inverse Langevin function with $\mathscr{L}(x)=\coth(x)-\dfrac{1}{x}$. Here, we employed a simple saturation-type evolution rule for $\lambda_{L}$ to capture the Mullins' effect.
\begin{equation}
\lambda_{L} = \lambda_{L, ss} - (\lambda_{L, ss} - \lambda_{L, 0})\exp{\left(-A_{\mathrm{N}}\left(\bar{\lambda}_{\mathrm{N},\max} - 1 \right)\right)}
\label{eq:lambdaLevolution}
\end{equation}
where $A_{\mathrm{N}}$ is a rate parameter, $\lambda_{L, 0}$ is the initial limiting chain extensibility, $\lambda_{L,ss}$ is the saturated limiting chain extensibility and $\bar{\lambda}_{\mathrm{N},\max}$ is the maximum average stretch over the loading history. Furthermore, we follow a constraint for the chain network alteration, $\mu_{\mathrm{N},0} \lambda_{\mathrm{N},0}^{2} = \mu_{\mathrm{N}} \lambda_{\mathrm{N}}^{2}$ where  $\mu_{0,\mathrm{N}}$ is the initial shear modulus. The constitutive equations and material parameters for the hard component are listed in Table \ref{Tab:material_parameter}.

\begin{table}[h]
\doublespacing
\centering
\small{
\begin{tabular}{lcc}
\hline
\textbf{Time-dependent elastic-inelastic mechanism I} &  &  \\
\hline
$\mathbf{T}_{\mathrm{I}} = \frac{1}{J} \mathbf{R}^{e}_{\mathrm{I}} \left( 2\mu_{\mathrm{I}} (\ln \mathbf{U}^{e}_{\mathrm{I}})_{0} \, + \, K(\ln J_{\mathrm{I}})\mathbf{I} \right) \mathbf{R}^{e \top}_{\mathrm{I}}$ & $\mu_{\mathrm{I}}$ [MPa] & 180.7 \\
& $K$ [GPa] & 1.75 \\
$\dot{\gamma}^{p}_{\mathrm{I}} = \dot{\gamma}_{\mathrm{I},0} \exp \left(- \frac{\Delta G_{\mathrm{I}}}{k_{B} \theta} \right) \sinh \left(\frac{\Delta G_{\mathrm{I}}}{k_{B} \theta} \frac{{\tau}_{\mathrm{I}}}{s_{\mathrm{I}}} \right) $ & $\Delta G_{\mathrm{I}}$ [$10^{-20}$J] & 2.9 \\
& $\dot{\gamma}_{\mathrm{I},0}$ [s$^{-1}$] & 0.028 \\
$\dot{s}_{\mathrm{I}} = h_{\mathrm{I}} \left(1-\frac{s_{\mathrm{I}}}{s_{\mathrm{I},ss}} \right)\dot{\gamma}^{p}_{\mathrm{I}}$ & $s_{\mathrm{I},ss}$ [MPa] & 0.55$s_{\mathrm{I},0}$ \\
& $h_{\mathrm{I}}$ [MPa] & 30 \\
\hline
\textbf{Time-independent hyperelastic network N} &  &   \\
\hline
$\mathbf{T}_{\mathrm{N}} = \frac{\mu_{\mathrm{N}}}{3J} \frac{\lambda_{L}}{\bar{\lambda}_{\mathrm{N}}} \mathscr{L}^{-1} \left(\frac{\bar{\lambda}_{\mathrm{N}}}{\lambda_{L}}\right) (\bar{\mathbf{B}}_{\mathrm{N}})_{0}$ & $\mu_{\mathrm{N},0}$ [MPa] & 29.67  \\
& $\lambda_{L,0}$ & $\sqrt5$ \\
$\lambda_{L} = \lambda_{L,ss} - (\lambda_{L,ss} - \lambda_{L,0})\exp{\left(-A_{\mathrm{N}}\left(\bar{\lambda}_{\mathrm{N},\max} - 1 \right)\right)}$ & $\lambda_{L,ss}$ & 1.6$\lambda_{L,0}$ \\
& $A_{\mathrm{N}}$ & 10 \\
\hline
\end{tabular}
}
\caption{Material parameters used in the constitutive model for the hard component.}
\label{Tab:material_parameter}
\end{table}

A nearly incompressible hyperelastic neo-Hookean representation (with the initial Poisson’s ratio $\sim$ 0.493) was used for the soft component, where the initial elastic modulus was taken to be $\sim$ 1/100 or $\sim$ 1/50 of the initial modulus of the hard component (\cite{lempesis2017atomistic}).
\begin{equation}
\mathbf{T}_{\mathrm{soft}} = J^{-1} \left( \mu_{\mathrm{soft}} \bar{\mathbf{B}}_{0} + K_{\mathrm{soft}}(\ln J)\mathbf{I} \right)
\end{equation}
where $\mu_{\mathrm{soft}}=2$ MPa and $K_{\mathrm{soft}}=142.5$ MPa for the stiffness ratio of  $E_{\mathrm{soft}}/E_{\mathrm{hard}} = 1/100$ and $\mu_{\mathrm{soft}}=4$ MPa and $K_{\mathrm{soft}}=285$ MPa for $E_{\mathrm{soft}}/E_{\mathrm{hard}} = 1/50$. An Arruda-Boyce eight chain representation was also tested for the soft domain with a large value of the limiting chain extensibility. The micromechanical modeling results with the Arruda-Boyce model were in turn found to be very similar to those with the neo-Hookean model. 

\begin{figure}[h]
\centering
\includegraphics[width=0.65\textwidth]{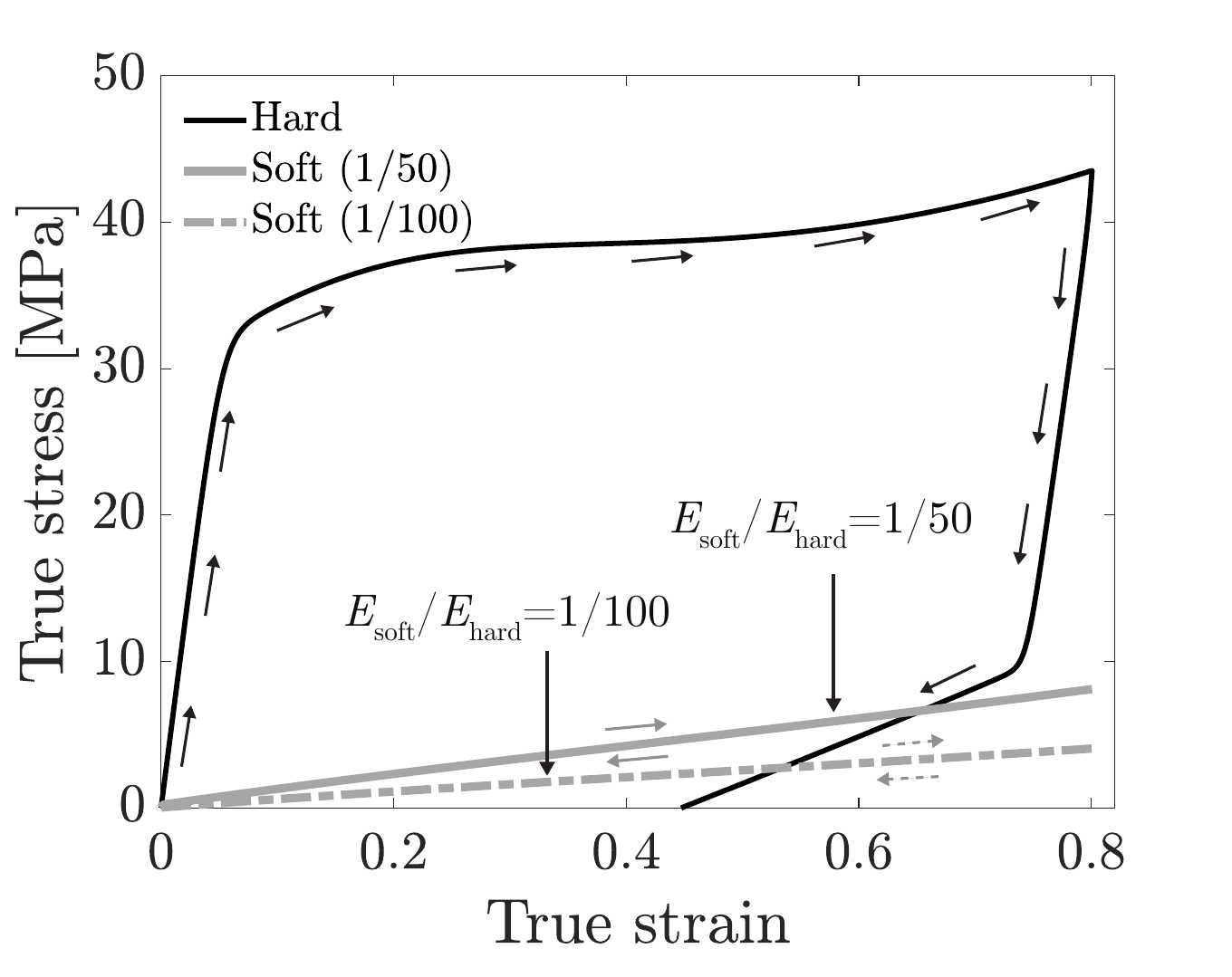} 
\caption{Stress-strain curves of hard and soft components used in the micromechanical models presented in Section \ref{section4}. Material parameters are given in Table \ref{Tab:material_parameter}.}
\label{fig:material}
\end{figure}

The stress-strain curves for the hard and soft components in loading and unloading at a strain rate of 0.01 $\mathrm{s}^{-1}$ are presented in Figure \ref{fig:material}. The hard component exhibited a thermoplastic-like behavior with significant energy dissipation and residual strain upon unloading while the soft component exhibited a hyperelastic behavior with no hysteresis and residual strain.

\subsection{Numerical implementation of the constitutive models for use in micromechanical analysis}
We then implemented a semi-implicit time integration procedure for constitutive equations, internal variables and the relevant finite deformation kinematic tensors, especially for the hard component summarized in Table \ref{Tab:material_parameter}, for use in finite element computations of the nonlinear boundary value problems for the RVEs presented in the main body of this paper. The time integration procedures were adopted in large part from \cite{weber1990finite} and \cite{chester2011mechanics}, which have been widely utilized for isotropic hyperelastic-viscoplastic materials. Here, the time integration procedure is briefly outlined especially for the elastic-inelastic micro-rheological mechanism in the hard component. The numerical implementation of the hyperelastic network mechanism N in the hard component is relatively straightforward since it is not time-dependent. Moreover, we have used a closed-form for the evolution of hyperelastic network modulus for the Mullins’ effect in Equation (\ref{eq:lambdaLevolution}). 

We summarize the time integration procedure as follows:
\begin{itemize}
\item Given : $\{\mathbf{F}(t), \;\mathbf{F}(\tau), \;\mathbf{F}^{p}_{\mathrm{I}}(t), \; \mathbf{T}_{\mathrm{I}}(t), \; s_{\mathrm{I}}(t), \; \dot{\gamma}^{p}_{\mathrm{I}}(t)\}$ at time $t$
\item Calculate : $\{\mathbf{T}_{\mathrm{I}}(\tau), \; s_{\mathrm{I}}(\tau), \; \dot{\gamma}^{p}_{\mathrm{I}}(\tau)\}$ at time $\tau = t + \Delta t$
\end{itemize}

\noindent Using an exponential map for integrating $\dot{\mathbf{F}}^{p}_{\mathrm{I}}=\mathbf{D}^{p}_{\mathrm{I}}\mathbf{F}^{p}_{\mathrm{I}}$, the inelastic deformation gradient is calculated,
\begin{equation}
\mathbf{F}^{p}_{\mathrm{I}}(\tau) =\exp\left(\Delta t\mathbf{D}^{p}_{\mathrm{I}}(\tau)\right)\mathbf{F}^{p}_{\mathrm{I}}(t).
\end{equation}
\noindent Then, the elastic deformation gradient is obtained, 
\begin{equation}
\begin{aligned}
\mathbf{F}^{e}_{\mathrm{I}}(\tau)  & = \mathbf{F}(\tau)\mathbf{F}^{p-1}_{\mathrm{I}}(\tau) \\
& =\mathbf{F}(\tau)\mathbf{F}^{p-1}_{\mathrm{I}}(t)\exp\left(-\Delta t\mathbf{D}^{p}_{\mathrm{I}}(\tau)\right) \\
& =\mathbf{F}^{e}_{\text{trial}}\exp\left(-\Delta t\mathbf{D}^{p}_{\mathrm{I}}(\tau)\right) \qquad \qquad \text{where} \qquad \mathbf{F}^{e}_{\text{trial}} & \defeq\mathbf{F}(\tau)\mathbf{F}^{p-1}_{\mathrm{I}}(t).
\end{aligned}
\end{equation}

\noindent The polar decomposition of the trial elastic deformation gradient is given by,
\begin{equation}
\begin{aligned}
\mathbf{F}^{e}_{\text{trial}} & =\mathbf{R}^{e}_{\text{trial}}\mathbf{U}^{e}_{\text{trial}} \\
\mathbf{R}^{e}_{\text{trial}} & \approx \mathbf{R}^{e}_{\mathrm{I}}(\tau) \\
\mathbf{U}^{e}_{\text{trial}} & \approx \mathbf{U}^{e}_{\mathrm{I}}(\tau)\exp\left(\Delta t\mathbf{D}^{p}_{\mathrm{I}}(\tau)\right).
\end{aligned}
\end{equation}
\noindent We then obtain the logarithmic elastic strain,
\begin{equation}
\begin{aligned}
\mathbf{E}^{e}_{\mathrm{I}}(\tau) & = \ln\left(\mathbf{U}^{e}_{\text{trial}}\exp\left(-\Delta t\mathbf{D}^{p}_{\mathrm{I}}(\tau)\right)\right) \\
& = \mathbf{E}^{e}_{\text{trial}}-\Delta t\mathbf{D}^{p}_{\mathrm{I}}(\tau) \qquad \qquad \qquad \text{where} \qquad \mathbf{E}^{e}_{\text{trial}} & =\ln\mathbf{U}^{e}_{\text{trial}}.
\end{aligned}
\end{equation}

\noindent Here, the relationship between the trial Mandel stress and the Mandel stress at time $t_{\mathrm{I}}(\tau)$ is obtained using the linear elasticity with the logarithmic strain measure, 
\begin{equation}
\begin{aligned}
\mathbf{M}^{e}_{\mathrm{I}}(\tau) & =2\mu_{\mathrm{I}}\left(\mathbf{E}^{e}_{\mathrm{I}}(\tau)\right)_{0}+K\text{tr}(\mathbf{E}^{e}_{\mathrm{I}}(\tau))\mathbf{I} \\
& = 2\mu_{\mathrm{I}}\left(\mathbf{E}^{e}_{\text{trial}}\right)_{0} -2\mu_{\mathrm{I}}\Delta t\mathbf{D}^{p}_{\mathrm{I}}(\tau) + K\text{tr}(\mathbf{E}^{e}_{\text{trial}})\mathbf{I} \\
& = \mathbf{M}^{e}_{\text{trial}}-2\mu_{\mathrm{I}}\Delta t\mathbf{D}^{p}_{\mathrm{I}}(\tau) \\
\text{where} \qquad \mathbf{M}^{e}_{\text{trial}} & =2\mu_{\mathrm{I}}\left(\mathbf{E}^{e}_{\text{trial}}\right)_{0}+K\text{tr}(\mathbf{E}^{e}_{\text{trial}})\mathbf{I} \\
\end{aligned}
\end{equation}
using $\text{tr}(\mathbf{E}^{e}_{\mathrm{I}}(\tau))=\text{tr}(\mathbf{E}^{e}_{\text{trial}})$ since $\mathbf{D}^{p}_{\mathrm{I}}(\tau)$ is deviatoric. Then, we have 
\begin{equation}
\left(\mathbf{M}^{e}_{\mathrm{I}}(\tau)\right)_{0}  = \left(\mathbf{M}^{e}_{\text{trial}}\right)_{0}-2\mu_{\mathrm{I}}\Delta t\mathbf{D}^{p}_{\mathrm{I}}(\tau).
\end{equation}

\noindent By defining,
\begin{equation}
\begin{aligned}
& \mathbf{D}^{p}_{\mathrm{I}}(\tau) =\dot{\gamma}^{p}_{\mathrm{I}}(\tau)\mathbf{N}^{p}_{\mathrm{I}}(\tau) & \\
& \bar{\tau}_{\mathrm{I}}(\tau) =\frac{1}{\sqrt{2}}\left\|\left(\textbf{M}^{e}_{\mathrm{I}}(\tau)\right)_0\right\| \qquad & \text{and} \qquad  \mathbf{N}^{p}_{\mathrm{I}}(\tau)=\frac{\left(\mathbf{M}^{e}_{\mathrm{I}}(\tau)\right)_{0}}{\sqrt{2}\bar{\tau}_{\mathrm{I}}(\tau)}\\
& \bar{\tau}_{\text{trial}} =\frac{1}{\sqrt{2}}\left\|\left(\textbf{M}^{e}_{\text{trial}}\right)_0\right\| \qquad & \text{and} \qquad  \mathbf{N}^{p}_{\text{trial}}=\frac{\left(\mathbf{M}^{e}_{\text{trial}}\right)_{0}}{\sqrt{2}\bar{\tau}_{\text{trial}}}, \\
\end{aligned}
\end{equation}
we obtain 
\begin{equation}
\bar{\tau}_{\mathrm{I}}(\tau)\mathbf{N}^{p}_{\mathrm{I}}(\tau)-\bar{\tau}_{\text{trial}}\mathbf{N}^{p}_{\text{trial}}+\sqrt{2}\mu_{\mathrm{I}}\Delta t\dot{\gamma}^{p}_{\mathrm{I}}(\tau)\mathbf{N}^{p}_{\mathrm{I}}(\tau)=0.
\end{equation}

\noindent Consequently, the following important results are given by,
\begin{equation}
\begin{aligned}
& \mathbf{N}^{p}_{\mathrm{I}}(\tau)=\mathbf{N}^{p}_{\text{trial}} \\
& \bar{\tau}_{\mathrm{I}}(\tau)-\bar{\tau}_{\text{trial}}+\sqrt{2}\mu_{\mathrm{I}}\Delta t\dot{\gamma}^{p}_{\mathrm{I}}(\tau)=0.
\end{aligned}
\end{equation} 

\noindent We solve the implicit equation for $\dot{\gamma}^{p}_{\mathrm{I}}(\tau)$ with the flow model from Equation (\ref{eq:gdflowrule}),
\begin{equation}
s_{\mathrm{I}}(\tau)\left(\frac{k_{B}\theta}{\Delta G_{\mathrm{I}}}\ln\frac{\dot{\gamma}^{p}_{\mathrm{I}}(\tau)}{\dot{\gamma}_{0}}+1\right)-\bar{\tau}_{\text{trial}}+\sqrt{2}\mu_{\mathrm{I}}\Delta t\dot{\gamma}^{p}_{\mathrm{I}}(\tau)=0.  \\
\end{equation} 
Here, $s_{\mathrm{I}}(\tau)$ is obtained using the Euler backward time integration,
\begin{equation}
s_{\mathrm{I}}(\tau) = s_{\mathrm{I}}(t) + \Delta t\dot{s}_{\mathrm{I}}(\tau) \qquad \text{where} \qquad \dot{s}_{\mathrm{I}}(\tau)=h_{\mathrm{I}}\left(1-\frac{s_{\mathrm{I}}(\tau)}{s_{ss, \mathrm{I}}}\right)\dot{\gamma}^{p}_{\mathrm{I}}(\tau) \\
\end{equation}
Then
\begin{equation}
s_{\mathrm{I}}(\tau) = \frac{s_{\mathrm{I}}(t)+\Delta t \, h_{\mathrm{I}} \, \dot{\gamma}^{p}_{\mathrm{I}}(\tau)}{1+\Delta t \, h_{\mathrm{I}}\, \dot{\gamma}^{p}_{\mathrm{I}}(\tau)/s_{ss, \mathrm{I}}}.
\end{equation}
The implicit equation for $\dot{\gamma}^{p}_{\mathrm{I}}(\tau)$ is rewritten in,
\begin{equation}
\left(\frac{s_{\mathrm{I}}(t)+\Delta t \, h_{\mathrm{I}} \,\dot{\gamma}^{p}_{\mathrm{I}}(\tau)}{1+\Delta t \, h_{\mathrm{I}} \,\dot{\gamma}^{p}_{\mathrm{I}}(\tau)/s_{ss, \mathrm{I}}}\right)\left(\frac{k_{B}\theta}{\Delta G_{\mathrm{I}}}\ln\frac{\dot{\gamma}^{p}_{\mathrm{I}}(\tau)}{\dot{\gamma}_{0}}+1\right)-\frac{1}{\sqrt{2}}\left\|2\mu_{\mathrm{I}}\left(\mathbf{E}^{e}_{\text{trial}}\right)_{0}\right\|+\sqrt{2}\mu_{\mathrm{I}}\Delta t\dot{\gamma}^{p}_{\mathrm{I}}(\tau)=0.
\label{eq:nonlinearimp}
\end{equation} 

\noindent The implicit equation in Equation (\ref{eq:nonlinearimp}) is then solved using a combination of bisection and Newton algorithms in \cite{NumericalRecipe}. We finally obtain the inelastic deformation gradient, $\mathbf{F}^{p}_{\mathrm{I}}(\tau)$ at time $\tau$ and hence the Cauchy stress $\mathbf{T}_{\mathrm{I}}(\tau)$ with the elastic deformation gradient $\mathbf{F}^{e}_{\mathrm{I}}(\tau)=\mathbf{F}(\tau)\mathbf{F}^{p-1}_{\mathrm{I}}(\tau)$ at time $\tau$.

The tangent moduli also known as the fourth-order Jacobian tensor consistent with the constitutive model were numerically computed via perturbation as documented well in \cite{kalidindi1992crystallographic} and \cite{chester2011mechanics}. The consistent elastic-plastic tangent moduli were then used for a Newton-type iterative procedure in searching for a solution that satisfies the global equilibrium in a weak manner for the boundary value problems for the RVEs.

\section{Finite element discretization and meshing strategies}
\label{appendixB}
The RVEs presented in Section \ref{section3} and Section \ref{section4} were systematically discretized for finite element computations. More specifically, mirroring and subsequent meshing strategies from edges, surfaces to volumes were carefully established such that nodes at pairs of periodic corners, edges and surfaces match perfectly, by which the kinematic constraints for the three-dimensional boundary conditions (Equations (\ref{eq:pbc}) – (\ref{eq:macroCauchystress})) were properly imposed to the RVEs. All of the meshing for the RVEs have been conducted in an advanced meshing program, \cite{CoreformCubit}. Furthermore, we used quadratic elements for both hard and soft domains in all of the RVEs used in our micromechanical analysis.

\section{Detailed geometric information for representative volume elements used in micromechanical analysis}
\label{appendixC}
% In the Supporting Information to this paper, we provide geometric details (coordinates for all points in the 3 by 3 by 3 unit-cells and neighbor lists for the central 7 Voronoi points) for an exemplar RVE ($N=7$) used in our micromechanical analysis: TPU$_{\mathrm{C}}$\_cont with v$_{\mathrm{hard}}=52.2 \%$ in Figure \ref{fig:cycle}.
In the Supporting Information to this paper, we provide geometric details (coordinates for all points in the 3 by 3 by 3 unit-cells and neighbor lists for the central Voronoi points) for exemplar RVEs ($N=7$) used in our micromechanical analysis: TPU$_{\mathrm{A}}$\_cont with v$_{\mathrm{hard}}=26.9 \%$ (Figure \ref{fig:tpuA_load_unload}), TPU$_{\mathrm{C}}$\_cont with v$_{\mathrm{hard}}=52.2 \%$ (Figure \ref{fig:tpuC_load_unload}, \ref{fig:cycle}, \ref{fig:mullins}) and TPU$_{\mathrm{C}}$\_cont with v$_{\mathrm{hard}}=50 \%$ (Figure \ref{fig:soft_in_hard}).

\end{appendices}

\clearpage
\printbibliography
\end{document}